\documentclass[twocolumn]{autart}    % Enable this line and disable the 
\usepackage{graphicx}          % Include this line if your 
\usepackage{cite}
\usepackage[round, sort]{natbib}
\usepackage{amsmath,amssymb,amsfonts}
\usepackage{algorithmic}
\usepackage{graphicx}
\usepackage{textcomp}
\usepackage{color}
\usepackage{xcolor}
\usepackage{amstext}
\usepackage{booktabs}
\usepackage{stfloats}
\usepackage{subfigure}
\newtheorem{remark}{Remark}{}
\newtheorem{theorem}{Theorem}

\newtheorem{definition}{Definition}

\newtheorem{corollary}{Corollary}
\usepackage{mathrsfs}
\usepackage{diagbox}
\allowdisplaybreaks[4]
\usepackage{cases}
\usepackage{xcolor}

\usepackage{xpatch}

\makeatletter
\ExplSyntaxOn
\cs_new:Npn \bibColoredItems #1#2
{
	\clist_map_inline:nn {#2} { \cs_new:cpn {bib@colored@##1} {#1} } 
}
\ExplSyntaxOff

\newcommand\bib@setcolor[1]{%
	\ifcsname bib@colored@#1\endcsname
	\expandafter\color\expandafter{\csname bib@colored@#1\endcsname}
	\else
	\normalcolor
	\fi
}

\xpatchcmd\@bibitem
{\item}
{\bib@setcolor{#1}\item}
{}{\PatchFailed}

\xpatchcmd\@lbibitem
{\item}
{\bib@setcolor{#2}\item}
{}{\PatchFailed}
\makeatother

\begin{document}

\begin{frontmatter}

\title{Resilient control of networked switched systems subject to deception attack and DoS attack} 

\thanks[footnoteinfo]{
	This work was supported by the National Natural Science Foundation of China (No. 62173243  and No. 61933014) and the Foundation of Key Laboratory of
	System Control and Information Processing, Ministry of Education, Shanghai,
	200240. The material in this paper was not presented at any conference. Corresponding author Z.  Zuo. Tel. +86-22-27890729.
	Fax +86-22-27406130.}
\author[c]{Rui Zhao}\ead{ruizhao@tju.edu.cn},    % Add the 
\author[c]{Zhiqiang Zuo$^*$}\ead{zqzuo@tju.edu.cn},               % e-mail address 
\author[a]{Ying Tan} \ead{ yingt@unimelb.edu.au},
\author[c]{Yijing Wang}\ead{yjwang@tju.edu.cn} , % (ead) as shown
\author[b]{Wentao Zhang}\ead{wentao.zhang@ntu.edu.sg}

\address[c]{Tianjin Key Laboratory of Intelligent Unmanned Swarm Technology and System, School of Electrical and Information Engineering, Tianjin University, Tianjin 300072, China.}  % Please
\address[a] {Department of Mechanical Engineering, University of Melbourne, VIC 3010, Australia.}
\address[b]{School of Electrical and Electronic Engineering, Nanyang Technological University, Singapore 639798. }


\begin{abstract}                          
In this paper, the resilient control for switched systems in the presence of deception attack and denial-of-service (DoS) attack is addressed. Due to the interaction of two kinds of attacks and the asynchronous phenomenon of controller mode and subsystem mode, the system dynamics becomes much more complex. A criterion is derived to ensure the mean square security level of the closed-loop system. This in turn reveals the balance of system resilience and control performance. 
\textcolor{black}{Furthermore, a mixed-switching control strategy is put forward to make the system globally asymptotically stable. It is shown that the system  will still converge to the equilibrium even if the deception attack occurs.}
Finally, simulations are carried out to verify the effectiveness of the theoretical results.
\end{abstract}

\end{frontmatter}
\section{Introduction}
A switched system is composed of continuous or discrete subsystems and a switching signal governing whether the underlying subsystem is activated or not.
It has been widely used in modeling complex engineering systems such as automobile transmission systems (\cite{App1}), power systems (\cite{App2}), intelligent traffic control systems (\cite{App3}) and so on. The main merits of switched systems can be summarized as follows \textcolor{black}{(\cite{ss}): 1) They can be used to model systems that are subject to known or unknown abrupt parameter variations which exhibit better control performance than the non-switched counterpart; 2) They can provide an effective and powerful mechanism to cope with highly complex systems and/or systems with large uncertainties which solve some control issues that are not amenable for a single controller in general.}
 Due to these advantages, much effort has been devoted to switched systems; see \cite{DT,DT3,Lin} and the references therein.

With the development of computer and network technology, wireless networks universally operate in the actual engineering (\cite{PSS}), for example, power systems, internet of vehicles, and other control processes. 
Of significant benefit is the reduction on the constraints of  physical distance of hardware devices (\cite{networked}).
As a result, the security issue naturally arises, which drastically degrades the control performance, and even leads to instability. 
Nowadays, cyber attacks have attracted considerable attention, and massive cyber attacks are exposed over time and scale. Typical examples include but are not limited to denial-of-service (DoS) attack (\cite{Feng2017}), deception attack (\cite{Wang2021}), replay attack (\cite{replay}), etc. The cyber attacks cause economic losses such as the Stuxnet worm attack on Iranian nuclear facilities (\cite{App-Dec}) and the Ukrainian power grid (\cite{App-com}), and so forth.

In order to defend against cyber attacks, non-switched systems subject to deception attacks  and DoS attacks have been considered in \cite{Zhao2020,guo2020}.
It should also be pointed out that the effort on the design of switched system has been made for secure analysis and synthesis; see \cite{Peng2020,Zhu2020} and the references therein.
In fact, the study on security issues for switched systems is an arduous task. 
On the one hand, the dynamic evolution becomes more complex for switched systems under attack in contrast to non-switched counterpart. On the other hand, the switching signal may be incompatible with the attacked systems, causing the degradation of the control performance. Actually, switched systems are more severely affected by the attacks; see the stabilization of linear switched systems under DoS attacks (\cite{DoS-SS1,Zhao2019,Qu2020}) and deception attacks  (\cite{Li2021,Yang2020,Qi2021,Gong2020}).

An approach to cyber attack in dynamic systems has been poured into the topic \textcolor{black}{for systems under deception attack and DoS attack}.
In general, it should be emphasized that the results of a single type of attack cannot be directly extended to the composite ones.
Actually, much progress has been made against multiple attacks; see \cite{IET2016}. 
In \cite{hybrid1,hybrid2,hybrid3}, a DoS attack restricted by the attack frequency or attack duration was put forward, and the deception attack obeys the Bernoulli distribution.  However, these ideas may not be immediately applicable to handle the stabilization of networked switched systems under \textcolor{black}{deception attack and DoS attack}. 

\textcolor{black}{There are few papers focusing on the resilience for switched system under DoS attack and deception attack.} 
\cite{han2022} studied the \textcolor{black}{deception attack and DoS attack} for switched system. Unfortunately, this paper does not consider the asynchronous behavior caused by DoS attack and the case where the bound of deception attack is not related to the current state. 
\textcolor{black}{As pointed out in \cite{asy1}, DoS attacks may cause asynchronous switching,	resulting in system performance degradation or even instability.  
More specifically,  the occurrence of asynchronous behavior has a close relationship with DoS attacks. Hence, the occurrence of asynchronous behavior is stochastic since DoS attack is stochastic. To the best of our knowledge, there is no method to deal with stochastic asynchronous behavior. A new approach is needed to design a switching signal to counter the negative influence of asynchronous behavior. }

\textcolor{black}{
In addition, only practical stability can be guaranteed in the existing results when the deception attack is involved, and little attention focuses on the issue of how to mitigate or eliminate bad effects caused by deception attacks. Furthermore, the advantage of switching signal is that the closed-loop dynamics can be stabilized even all subsystems are unstable. Unlike non-switched systems, 
the stability of switched systems is not dependent on controller but also the switching law, as pointed out in  \cite{ss}. Comparing with the results for the non-switched system, the switching signal provides another free variable for the switched system, which is an additional possibility to improve system resilience. 
Consequently, a problem arises: How to design an appropriate switching signal and a control strategy for the switched system to mitigate or eliminate the negative effect of deception attacks? }

To the best of the authors' knowledge, little work has been done on the stabilization of switched systems under \textcolor{black}{deception attack and DoS attack}. Notice that such an issue has theoretical challenges and engineering significance which motivates our current study.
The main contributions of this paper are:
{\color{black}
\begin{enumerate}
	\renewcommand{\labelenumi}{\theenumi)}
		\item The stabilization problem for switched systems under both DoS attack and deception attack is investigated via time-dependent switching law. It is shown that the security level is related to the initial state and the attack parameters. The controller gain is then designed to guarantee the $\ell$-security level for the closed-loop system.
\item \textcolor{black}{The relationship between the attack parameters and the upper bound of Lyapunov function is revealed, where the asynchronous behavior is caused by DoS attack. This suggests that an increased attack probability of DoS attack may lead to a larger ultimate bound with stronger requirements on the switching signals.} The connection between the ultimate bound and the attack parameters is established. This suggests that a fiercer attack leads to a larger asymptotic bound and a higher security level.
\item A mixed-switching control strategy is presented by incorporating both dwell time approach and state-dependent switching law. \textcolor{black}{With  such a switching signal, the system is asymptotically stabilizable even if the deception attack exists.}
\end{enumerate}}

The remainder of this paper is organized as follows. Section \ref{sec_pf} gives some preliminaries and system specification. Section \ref{sec_main} develops the main results on stabilization for switched systems under  DoS attack and deception attack. Numerical simulations are conducted in Section \ref{sec_sim} to support the derived theoretical analysis. Finally, we conclude this paper in Section \ref{sec_col}.

Notations:
Denote $\mathbb{R}$ the real number set, $\mathbb{N}$ the integer set, $\mathbb{N}_{\geq k}$ (resp. $\mathbb{R}_{\geq k}$) the integer (resp. real) number set no less than $k$.
The maximum (resp. minimum) eigenvalue of a square matrix is represented by $\lambda_{max}$ (resp. $\lambda_{min}$). 
$A^T$ stands for the transpose of matrix $A$. $P\succ0$ ($P\succeq 0$) represents a positive definite (semi-positive definite) matrix. $\textbf{0}$ and $\textbf{I}$ stand for zero matrix and unit matrix with appropriate dimensions. 

\section{Problem formulation}\label{sec_pf}

\begin{figure}[t]
	\centering
	\includegraphics[width=0.7\linewidth]{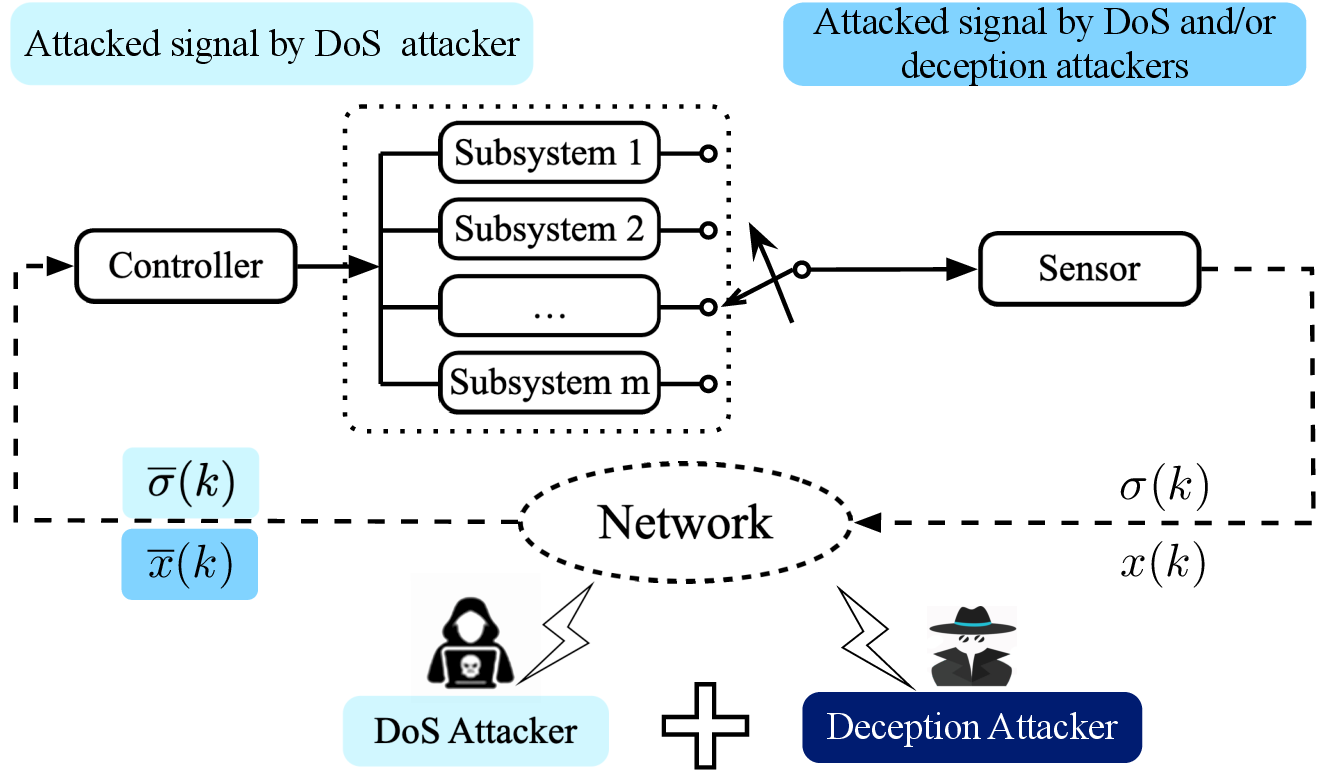}%{111111111.eps}%{Timing.eps}
	\caption{A framework for switched system under \textcolor{black}{deception attack and DoS attack}.}
	\label{Fig_closed}
\end{figure}
Consider a class of switched linear systems
\begin{equation}\label{state}
	\begin{aligned}
		x(k+1) = A_{\sigma(k)}x(k)+B_{\sigma(k)}u(k) 
	\end{aligned} 
\end{equation}
where $x(k)\in\mathbb{R}^{n_x}$ and $u(k)\in\mathbb{R}^{n_u}$ are state vector and control signal, respectively. \textcolor{black}{$\sigma(k):~\mathbb{N} \mapsto \mathcal{M}\triangleq\{1,2, \cdots, m\}$} is the switching signal with $m \in \mathbb{N}$ being the number of subsystems. $A_p, ~B_p$ ($p\in \mathcal{M}$) are constant matrices with compatible dimensions. The controller is with the form 
$	u(k) = K_{{\sigma}(k)}{x}(k)$,
where $K_{{\sigma}(k)}$ is the controller gain associated with the system mode.

The state information $x(k)$ and the switching signal $\sigma(k)$ are transmitted via the network, which is accessible for cyber attackers, see Fig. \ref{Fig_closed} for more details.  In this paper, both DoS attack and deception attack act on the sensor-to-controller channel. Once the attacks occur, the state signal and the switching signal received by controller become attacked state $\overline{x}(k)$ and attacked switching signal $\overline{\sigma}(k)$, whose detailed forms will be clear in the rest paper. Then the controller output suffering from attack is rewritten as $	u(k) = K_{\overline{\sigma}(k)}\overline{x}(k).$

Suppose that DoS attack and deception attack can occur at the same time. 
Next, we analyze the value $\overline{x}(k)$ when DoS attack and deception attack are individually or simultaneously  launched.

	\textbf{Deception attack:} The deception attack is usually realized by tampering with system data or packets, e.g., the attacker sends a false data packet to the target directly or deciphers the authenticated packet and injects the false data into the original packet; see \cite{def}.
	When the system is suffering deception attack, the measurement signal changes to $x_a(k)$, which satisfies
	\begin{equation}\label{deception_ass}
		\|x_a(k)\|_{\textcolor{black}{2}}\leq \overline{\gamma}^2,
	\end{equation} 
	with $\overline{\gamma}$ being a known positive constant.
	Therefore, the controller input becomes  $	\overline{x}(k)  =x_a(k)$ if the deception attack is successfully launched, i.e., $\alpha(k) =1$. Moreover, the deception attack obeys the Bernoulli distribution. $\alpha(k)$ is a binary random variable characterizing the occurrence of deception attacks. That is, $\alpha (k) =1$ when the attack occurs, and $\alpha (k) = 0$ otherwise. And their probabilities are
$
		\text{Prob}\{\alpha(k) = 1\} = \overline{\alpha},~\text{Prob}\{\alpha(k) = 0\} =1- \overline{\alpha},
$
	where $\overline{\alpha} \in [0,1]$ is a known constant.

	\textbf{DoS attack:} The DoS attack attempts to temporarily or permanently block the service of devices connected to the network, thereby rendering network resources unavailable to legitimate users, see \cite{def}.
	When the system is subject to DoS attack, the measurement signal and the switching signal will be lost. The controller input turns to be the latest one, i.e., $	\overline{x}(k)  = \overline x(k-1)$. Similarly, the DoS attack also has some effect on the switching signal, that is, the controller mode admits $\overline{\sigma}(k)  = \overline \sigma(k-1)$ if there exists a DoS attack.
	The DoS  attack also follows the  Bernoulli distribution, i.e., $\beta(k)$ is a binary random variable. When an attack occurs, $\beta (k) =1$, otherwise, $\beta (k) = 0$. Similarly, $\text{Prob}\{\beta(k) = 1\} = \overline{\beta},~\text{Prob}\{\beta(k) = 0\} = 1-\overline{\beta}$, where $\overline{\beta} \in [0,1]$ is a positive scalar and pre-known information for designer.

	\textbf{Simultaneous  attacks:} When the system is suffering two types of attacks simultaneously, i.e., $\alpha(k) =1$ and $\beta(k) =1$. However, no matter what attack signals are injected by deception attacker, they cannot be transmitted to the controller when the DoS attack is occurring. In  other words, the attack effect of \textcolor{black}{deception attack and DoS attack} is the same as that for single DoS attack.  Then the controller input becomes $\overline{x}(k)  = \overline x(k-1)$.
	{\color{black}
	\begin{remark}
		In the problem setting, both DoS attack and deception attack are assumed to satisfy Bernoulli distributions with known parameters. It is recognized that random attacks align more closely with engineering practice, capturing the complexities, randomness, and unpredictability inherent in network control systems, as emphasized in \cite{AE-6,AE-7}. Additionally, the effectiveness of attacks often depends on dynamic network conditions, such as network load, congestion, and transmission, as outlined in \cite{AE-8}. Given the inherently stochastic nature of these processes, describing attacks through random processes is considered a fitting representation, effectively capturing the inherent randomness embedded in attack patterns. While acknowledging that assuming Bernoulli distributions for both attacks might be considered simplistic, it's important to note that they offer a direct means to capture randomness and unpredictability in network behaviors. Our future work will explore more realistic random distributions in subsequent research endeavors. Furthermore, obtaining precise knowledge of parameters in practical scenarios may pose challenges. However, it is essential to emphasize that if the probability distribution and attack power are known, statistical methods can be employed to estimate these parameters from collected measurements \cite{AE-1,AE-2,AE-3}. We argue that, under the assumption that such probability distributions are slowly time-varying, this approach could be viable. To counter potential parameter drifting, we propose the possibility of continuously re-triggering offline parameter identification techniques when the currently identified parameters fail to yield satisfactory results. These avenues will be explored further in our future work.
\end{remark}}

{\color{black}
		\begin{remark}
	The reasons of modeling attacks obeying the Bernoulli distribution are three-folds.
	First of all, whether the adversary successfully targets a network is related to network random factor such as network load, network congestion and network transmission rate. Second, the defense mechanism of a system such as firewalls and/or detectors in practical applications will reduce the attack rate. Finally, the adversary's energy is limited, and it generally does not operate continuously. Therefore, it is reasonable to assume that the attack follows a Bernoulli distribution; see \cite{IET2016,Zhao2020,wu2021,guo2020}.
\end{remark}
\begin{remark}
		For deception attack, the boundedness on signal $x_a(k)$ is required. In practice, arbitrarily unbounded attacks would consume a great deal of energy, which are  always impossible. 
	Hence, it makes sense to require that the false signals injected by adversaries are bounded as in \cite{Zhao2020}.
	Although the assumption of knowing the bound of the deception attack’s power seems restrictive,  we think that it is a reasonable assumption. Engineers usually can obtain very conservative estimation of such bounds; see \cite{AE-3,IET2016}. As pointed out in \cite{AE-5}, the bound can be estimated through statistical tests and specified by security requirements. A conservative bound will enhance the robustness of the proposed control design.
\end{remark}}

	Based on the above analysis, the actual input of the controller  is
	{\color{black}	\begin{equation}\label{input}
			\begin{aligned}
			\overline{x}(k) 
			=~& \left(1-\alpha(k)\right)\left(1-\beta(k)\right)x(k) + \beta(k) \overline{x}(k-1)\\ &+\alpha(k)\left(1-\beta(k)\right)x_a(k).
			\end{aligned}
		\end{equation}}
	From this equation, it is clear that $\overline{x}(k) = x(k)$ when $\alpha(k) = 0$ and $\beta(k) = 0$, which means that the system is attack-free at instant $k$. If only deception attack occurs, i.e., $\alpha(k) =1$ and $\beta(k) = 0$, the signal controller received becomes $\overline{x}(k) = x_a(k)$. In addition, when DoS attack is launched the controller input admits $\overline{x}(k) = x(k-1)$.

{\color{black}	
\begin{remark}
	In this paper, the scenario where the deception attack and DoS attack are independent is considered, which means that the DoS attack and deception attack can occur at the same time. The result can be extended to the case where DoS attack and deception attack are dependent. In such a case, DoS attack and deception attack cannot launch simultaneously. More specifically, there are three situations: 1) $\alpha(k)=1$, $\beta(k)=0$; 2) $\alpha(k)=0$, $\beta(k)=0$; 3) $\alpha(k)=0$, $\beta(k)=1$. To sum up, $\alpha(k) \beta (k) \equiv 0$. Hence, using the same method as that in (3), the controller input becomes $\overline{x}(k)
	=~ \left(1-\alpha(k)-\beta(k)\right)x(k) + \beta(k) \overline{x}(k-1)+\alpha(k)x_a(k)$. The following analysis is still valid for dependent attacks.  
	
\end{remark}}

As we can see, ${\overline{\sigma}(k)}$ is the controller mode described by $\overline{\sigma}(k)  =\beta(k) \overline {\sigma}(k-1) + (1-\beta(k)) \sigma(k) $ since the switching signal is only effected by DoS attack.
It is well known that the controller mode and the subsystem mode may be mismatching, i.e., $\overline{\sigma}(k)\neq \sigma(k)$, which is called asynchronous behavior; see \cite{Zhang}.  
 More importantly, the system performance may degrade  at the asynchronous stage since the mismatching controller cannot stabilize the activated subsystem.
 Hence, the dynamics of the closed-loop system is  divided into the synchronous stage ($\sigma(k) = \overline{\sigma}(k)$) and the asynchronous one ($\sigma(k) \neq \overline{\sigma}(k)$). Note that the system is in the asynchronous stage only when the DoS attack occurs.

According to relation of $\sigma(k)$ and $\overline{\sigma}(k)$, we give the closed-loop dynamics under attacks, respectively. 
For the synchronous stage, i.e., $\sigma(k) = \overline{\sigma}(k)$, the dynamics  of the closed-loop system is
\begin{align}
		&\widetilde{x}(k+1) \label{clo-sy1}\\
		= ~&\mathcal{A}_{\sigma(k)}^1\widetilde{x}(k) +(\beta(k)-\overline{\beta})\mathcal{A}_{\sigma(k)}^2 \widetilde{x}(k)- \chi_1 \mathcal{A}_{\sigma(k)}^3 \widetilde{x}(k) \notag\\
		&+ \chi_2 \mathcal{A}^4_{\sigma(k)}x_a(k)+\overline{\alpha}(1-\overline{\beta}) \mathcal{A}^4_{\sigma(k)}x_a(k)\notag
\end{align}
where $\widetilde{x}(k) = [x^T(k),\overline{x}^T(k-1)]^T$, $\chi_1 = \alpha(k)-\overline{\alpha} + \beta(k)-\overline{\beta} - \alpha(k)\beta(k)+\overline{\alpha}\overline{\beta}$, $\chi_2 =\alpha(k)-\overline{\alpha}-\alpha(k)\beta(k)+\overline{\alpha}\overline{\beta}$, $\chi_3 =(1-\overline{\alpha}-\overline{\beta}+\overline{\alpha}\overline{\beta})$,
\begin{align*}
	\mathcal{A}_{\sigma(k)}^1 &= \left[ 
	\begin{array}{cc}  
		A_{\sigma(k)}+\chi_3 B_{\sigma(k)}K_{\sigma(k)}&  \overline{\beta} B_{\sigma(k)}K_{\sigma(k)}\\  %第一行元素
		\chi_3 \textbf{I}& \overline{\beta} \textbf{I} %第二行元素
	\end{array}
	\right],\\
	\mathcal{A}_{\sigma(k)}^2 &= \left[ 
	\begin{array}{cc}  
		\textbf{0}& B_{\sigma(k)}K_{\sigma(k)}\\  %第一行元素
		\textbf{0}& \textbf{I} %第二行元素
	\end{array}
	\right],	\\
	\mathcal{A}_{\sigma(k)}^3 &= \left[ 
	\begin{array}{cc}  
		B_{\sigma(k)}K_{\sigma(k)} & \textbf{0} \\  %第一行元素
		\textbf{I} & \textbf{0} %第二行元素
	\end{array}
	\right],~	\mathcal{A}_{\sigma(k)}^4= \left[ 
	\begin{array}{c}  
		B_{\sigma(k)}K_{\sigma(k)} \\  %第一行元素
		\textbf{I}%第二行元素
	\end{array}
	\right].
\end{align*}

{ 	For the asynchronous stage, DoS attack is successfully launched, i.e., $\beta(k) = 1$. The closed-loop system over the asynchronous stage becomes
	\begin{equation}\label{asy-clo}
		\begin{aligned}
			&\widetilde{x}(k+1) = ~\overline{\mathcal{A}}_{\sigma(k),\overline{\sigma}(k)}^1\widetilde{x}(k)
		\end{aligned}
	\end{equation}
	where 
	$	\overline{\mathcal{A}}_{\sigma(k),\overline{\sigma}(k)}^1= \left[ 
	\begin{array}{cc}  
		A_{\sigma(k)}& B_{\sigma(k)}K_{\overline{\sigma}(k)}\\  %第一行元素
		\textbf{0} & \textbf{I} %第二行元素
	\end{array}
	\right]$.
	
	Formula (\ref{asy-clo}) only describes the evolutionary process from a switching instant to the moment at which the switching signal is successful sent. \textcolor{black}{(Switching instant is the moment at which $\sigma(k_s) \neq \sigma(k_s-1)$)}  Similar to (\ref{asy-clo}), the dynamic at the first successful transmitted instant after a switching is also especial.  At this instant, the system mode and the controller mode are matched, i.e., $\beta(k) = 0$. Hence, the closed-loop dynamics admits
\begin{equation}\label{clo-sy2}
	\begin{aligned}
			\widetilde{x}(k+1) =~&\widetilde{\mathcal{A}}^1_{\sigma(k)} \widetilde{x}(k) - (\alpha(k)-\overline{\alpha})  \widetilde{\mathcal{A}}^2_{\sigma(k)}  \widetilde{x}(k)\\
			&+\alpha(k) \widetilde{\mathcal{A}}^3_{\sigma(k)} x_a(k) 
	\end{aligned}
\end{equation}
where 
\begin{align*}
	\widetilde{\mathcal{A}}_{\sigma(k)}^1 &= \left[ 
	\begin{array}{cc}  
		A_{\sigma(k)} + (1- \overline{\alpha}) B_{\sigma(k)}K_{\sigma(k)}& \textbf{0}\\  %第一行元素
		(1- \overline{\alpha})\textbf{I} & \textbf{0} %第二行元素
	\end{array}
	\right],\\
	\widetilde{\mathcal{A}}_{\sigma(k)}^2 &= \left[ 
	\begin{array}{cc}  
		B_{\sigma(k)}K_{\sigma(k)} &\textbf{0}\\  %第一行元素
		\textbf{I}&\textbf{0}
	\end{array}
	\right],~
	\widetilde{\mathcal{A}}_{\sigma(k)}^3 = \left[ 
	\begin{array}{c}  
		B_{\sigma(k)}K_{\sigma(k)} \\  %第一行元素
		\textbf{I}
	\end{array}
	\right].
\end{align*}

{\color{black}\begin{remark}
		This paper considers the scenario where deception attack only changes the state signal. 
		Note that switching signals are well-designed by engineers or scholars, therefore they have a higher lever of security. In other words, it is more strictly protected and less vulnerable to attacks compared to system state. On the other hand, the presence of attacks on the switching signal will make the considered problem more complicated. This in turn imposes stronger constraints on deception attack, DoS attack and switching law.
		To the best of our knowledge, such an issue has not yet been addressed. 
\end{remark}}

In what follows, we give some definitions.

\begin{definition}[\hspace{-0.001cm}\cite{DT}]
	A positive constant $\tau_d\in \mathbb{R}$ is called the dwell time (DT) of a switching signal if for any $s\in \mathbb{N}$, \begin{equation}
		\tau_d \leq k_{s+1}-k_s\end{equation}
	where $k_s$ and $k_{s+1}$ are switching instants.
\end{definition}

\begin{definition}[\hspace{-0.001cm}\cite{def2}]\label{def2}
The switched system is said to be \textcolor{black}{practically stable}  in the mean square sense, if \textcolor{black}{for a given switching signal $\sigma(k)$}, there exist scalars $\varepsilon > 0$, $\eta\in [0,1)$, $\psi \geq 0$ such that 
	\begin{equation}\label{equ-ep}
		\mathbb{E} \{ \|\widetilde{x}(k)\|^2\} \leq \varepsilon \|x(0)\|^2 \eta^k + \psi, ~k \geq 0
	\end{equation}
where $\psi$ is the asymptotic bound.
\end{definition}

\begin{definition}[\hspace{-0.001cm}\cite{def3}]\label{def3}
	The switched system is said to be asymptotically stable  in the mean square sense, if
	\begin{equation}
\lim\limits_{k\rightarrow \infty} \mathbb{E} \{\|x(k)\|^2\}=0.
	\end{equation}
\end{definition}

	\begin{definition}[\hspace{-0.001cm}\cite{IET2016}]
		For a switched system suffering from attack with a given switching signal $\sigma(k)$, the following set
\begin{equation}
			\mathcal{R} = \left \{ \widetilde {x}_0 \in \mathbb{R}^{n_x}:	\mathbb{E}\left\{\|\widetilde{x}(k)\|^2\right\} \leq \ell, ~\forall k \geq 0 \right\}
	\end{equation}
		is called the mean square $\ell$-security set with $\ell>0$ being a desired security level.
\end{definition}

In this paper, we aim to guarantee that the switched system under DoS attack and deception attack is \textcolor{black}{practically stable} in the mean square sense by designing the controller gain and time-dependent switching signal, and then calculate the security set in the mean square. Moreover, a more flexible switching signal is suggested to ensure asymptotic stability in the mean square sense even if there exists deception attack.
{\color{black}Fig. \ref{fig:structure} illustrate the organization of this paper.
Theorem \ref{thm1} and Theorem \ref{thm2} deal with the stabilization problem  based on the time-dependent switching law. In Theorem \ref{thm2}, the controller is designed via  LMI conditions. Moreover, the mixed-switching law is devised in Theorem \ref{thm3} to finish asymptotic stability. Example 1 and Example 2 show the feasibility and superiority of Theorem 2 and 3, respectively. Example 3 illustrates that the results in this paper are suitable for practical application. }

\begin{figure}
	\centering
	\includegraphics[width=0.6\linewidth]{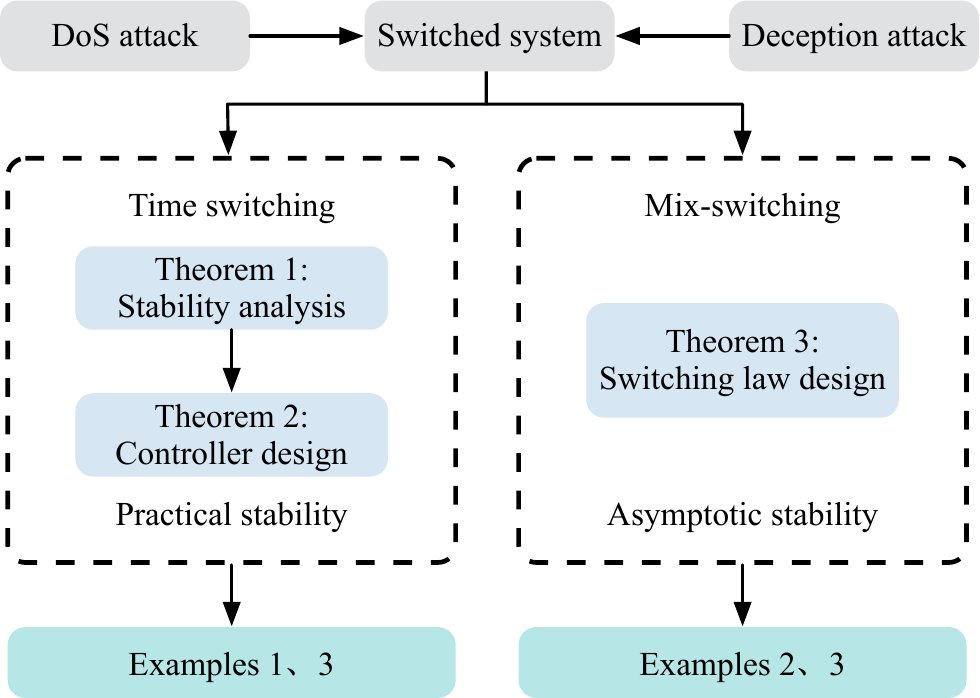}
	\caption{\textcolor{black}{The organization of this paper}}
	\label{fig:structure}
\end{figure}

\section{Main Results}\label{sec_main}

\subsection{Time-dependent switching signal}\label{subsec1}
In this subsection, we are dedicated to the stabilization issue for switched systems  subject to both DoS attack and deception attack. A sufficient condition is presented to ensure the \textcolor{black}{practical stability} of the  considered  systems in the mean square sense. 
\begin{theorem}\label{thm1}
	Given scalars $\mu > 1$, $0<  \rho_s <1 $, $\rho_u > 0$, $\overline{\alpha},~\overline{\beta} \in [0,1]$, $c= \overline{\beta}\frac{1+\rho_u}{1-\rho_s} < 1$, and matrices $K_p$ guaranteeing $A_p+B_pK_p$ are Schur stable. \textcolor{black}{If there exist matrices $P_p\succ0,~p\in \mathcal{M}$} and a positive constant $\epsilon$ such that
	\begin{align}
		{\Pi}_p= \left[ 
		\begin{array}{cc}  
			{\Pi}_p^{11}& {\Pi}_p^{12}\\
			*& {\Pi}_p^{22}
		\end{array}
		\right] \preceq 0, \label{con1}\\
		{\Omega}_{pq}=(\overline{\mathcal{A}}^1_{pq})^T P_p \overline{\mathcal{A}}^1_{pq} -  P_p - \rho_u P_p\preceq 0, \label{con2}\\
		{\Psi}_p= \left[ 
		\begin{array}{cc}  
			{\Psi}_p^{11}& {\Psi}_p^{12}\\
			*& {\Psi}_p^{22}
		\end{array}
		\right] \preceq 0, \label{con3}\\
		P_p \preceq \mu P_q,~~~~\textcolor{black}{\forall p,q \in \mathcal{M},~p\neq q}, \label{con4}
	\end{align}
	where $	\widetilde{\alpha} = ~\overline{\alpha}\left(1-\overline{\alpha}\right),~\widetilde{\beta} = \overline{\beta}\left(1-\overline{\beta}\right)$,~
	\begin{align*}
		\Pi_p^{11}	=~& (\mathcal{A}^1_p)^TP_p \mathcal{A}^1_p+\widetilde{\beta}(\mathcal{A}^2_p)^TP_p\mathcal{A}^2_p  - P_p + \rho_s P_p\\
			&+\left(\widetilde{\alpha}\left(1-\overline{\beta}\right)^2+\widetilde{\beta}\left(1-\overline{\alpha}\right)^2+
		\widetilde{\alpha}\widetilde{\beta}\right)(\mathcal{A}^3_p)^TP_p \mathcal{A}^3_p\\
		& +2\widetilde{\beta}\left(1-\overline{\alpha}\right)(\mathcal{A}^2_p)^TP_p\mathcal{A}^3_p,~\\
		\Pi_p^{12}=~&
		\overline{\alpha}\left(1-\overline{\beta}\right)(\mathcal{A}^1_p)^TP_p \mathcal{A}^4_p-\overline{\alpha}\widetilde{\beta} (\mathcal{A}^2_p)^TP_p \mathcal{A}^4_p\\&
		-\widetilde{\alpha}\left(1-\overline{\beta}\right)^2(\mathcal{A}^3_p)^TP_p \mathcal{A}^4_p,~\\
		\Pi_p^{22}	=~&	\overline{\alpha}\left(1-\overline{\beta}\right)(\mathcal{A}^4_p)^TP_p \mathcal{A}^4_p -\epsilon \textbf{I},~\\
		{\Psi}_p^{11} =~& (\widetilde{\mathcal{A}}^{1}_p)^T P_p \widetilde{\mathcal{A}}^{1}_p +\widetilde{\alpha}(\widetilde{\mathcal{A}}^{2}_p)^T P_p \widetilde{\mathcal{A}}^{2}_p -P_p+\rho_s P_p,~\\
		{\Psi}_p^{12} =~& \overline{\alpha} (\widetilde{\mathcal{A}}^{1}_p)^T P_p \widetilde{\mathcal{A}}^{3}_p -\widetilde{\alpha}(\widetilde{\mathcal{A}}^{2}_p)^T P_p \widetilde{\mathcal{A}}^{3}_p,~\\
		{\Psi}_p^{22} =~&  \widetilde{\alpha} (\widetilde{\mathcal{A}}_p^3)^T P_p \widetilde{\mathcal{A}}_p^3+\overline{\alpha}^2 (\widetilde{\mathcal{A}}_p^3)^T P_p \widetilde{\mathcal{A}}_p^3 - \epsilon \textbf{I}.
	\end{align*}
	Then switched system (\ref{state}) is \textcolor{black}{practically stable} in the mean square sense with $\ell$-security level, and the switching signal \textcolor{black}{satisfies} 
	\begin{equation}\label{taud}
		\tau_d \geq \tau_d^*=- \frac{\ln\overline{\mu}}{\ln\overline{\rho}_s}
	\end{equation}
	with $\overline{\mu} = \mu \frac{2-\overline{\beta}-c}{1-c}$, and\\
$\ell = \overline{\mu} \max \left\{ \textcolor{black}{\frac{\tilde{x}^T(0) P _{\sigma(0)} \tilde x(0)}{ \min_{p\in\mathcal{M}}\{\lambda_{\min}\left(P_p\right)\}},~\frac{\epsilon \overline{\gamma}^2}  {\min_{p\in\mathcal{M}}\{\lambda_{\min}\left(P_p\right)\}\rho_s}} \right\}$. 
	Moreover, the asymptotic bound of the switched system is 
$\psi =\frac{ \overline{\mu} \epsilon \overline{\gamma}^2  }{\rho_s\min_{p\in\mathcal{M}}\{\lambda_{\min}(P_p)\}}$.
\end{theorem}

\begin{pf}
	Please see Appendix \ref{app1}.\qed
\end{pf}

{\color{black}\begin{remark}
	Theorem \ref{thm1} gives a criterion that the switched system is \textcolor{black}{practically stable}  in the mean square sense under \textcolor{black}{deception attack and DoS attack}. The explicit value of the security level is formulated. More specifically, LMIs (\ref{con1}) and  (\ref{con2}) guarantee that the energy decrease or increase has an upper bound in the mean square sense if the system mode is the same as the controller mode when DoS attack occurs. On the other hand, LMI (\ref{con3}) makes sure that the Lyapunov function decreases for the deception attack. And LMI (\ref{con4}) limits a bounded energy jump at the switching instant.
\end{remark}}

\begin{remark}
	\textcolor{black}{	
		The results in this paper can be extended to the dynamic feedback controller. \cite{LMI} pointed out that the dynamic feedback controller allows us to meet more specifications than the static one. As illustrated in \cite{DC}, an augmented state can be constructed, which provides more design freedom to achieve better performance.}
\end{remark}

\begin{remark}{\color{black}
		 It is known that the increase of energy during asynchronous behavior is related to  the attack probability of DoS attack (\cite{asy2}).  With increasing number of successive attacks, the attack probability decreases.  Therefore, there is an upper bound for the energy growth  during asynchronous behavior. In this work, the upper bound calculated based on the DoS attack probability is used to characterize the maximum energy increase during asynchronous interval. Then a switching signal is designed to ensure system stability in the worst case scenario.}
		More specifically, 	the probability of consecutive DoS attacks decreases over time. Thus there exists a maximum incremental energy of asynchronous behavior caused by DoS attack,  which can be calculated by Theorem \ref{thm1}. The maximum energy increase in mean square sense is $\mu\frac{2-\overline{\beta}-c}{1-c}V(t_s)$ with  $c=\overline{\beta}\frac{1+\rho_u}{1-\rho_s}$ and $V(t_s)$ being the energy at switching instant $t_s$, which is associated with the system energy at the switching instant, the probability of DoS attacks, the synchronous convergence rate, and the asynchronous divergency rate. In Theorem \ref{thm1}, the energy increase is defined as $\overline{\mu}$.
		It is reasonable that a fiercer attack will cause undesirable energy divergence and be more difficult to control.
\end{remark}

{\color{black}	\begin{remark}
The LMI method is used to obtain sufficient conditions for switched systems under DoS attack and deception attack. The LMI method is conservative but also brings robustness. When the attack induced asynchronous behaviors are considered, the robustness of the proposed method will be further improved.
Moreover, the estimation of the deception attack bound is also conservative which increases robustness.  The existing robust control techniques can be incorporated into the proposed method to further enhance the robustness (\cite{AE-28,AE-29}).
\end{remark}}

\begin{remark}\label{rm_canshu}
	According to Theorem \ref{thm1}, the asymptotic bound is $\overline{\mu}\epsilon \overline{\gamma}^2\diagup (\rho_s \lambda_{\min}(P_p))$, which is equal to the security level subject to a sufficiently small value of  $\|\widetilde{x}(0)\|$. It is obvious that the asymptotic bound is related to the deception attack bound $\overline{\gamma}$, the DoS attack probability and deception attack probability. A larger probability of DoS attack and deception attack or the fierce deception level will yield a larger asymptotic bound. 
\end{remark}

 In contrast with the results in  \cite{han2022},  one challenge is to handle the  negative effect on the switching signal caused by DoS attack. Furthermore, the inequality constraints are related to the attack probability. Another difficulty compared with \cite{IET2016} is the complex dynamics of switched system due to multiple subsystems.

Next, some corollaries are given for the cases where  \textcolor{black}{deception attack and DoS attack} occur on non-switched system or single attack tampers switched system.

For non-switched system, i.e., $\mathcal{M} = \{1\}$, the dynamics of the closed-loop system boils down to 
\begin{equation*}
	\begin{aligned}
		\widetilde{x}(k+1) =& ~\mathcal{A}^1\widetilde{x}(k) +(\beta(k)-\overline{\beta})\mathcal{A}^2 \widetilde{x}(k)
		+\chi_1 \mathcal{A}^3 \widetilde{x}(k)\\&+\chi_2 \mathcal{A}^4x_a(k)+\overline{\alpha}(1-\overline{\beta}) \mathcal{A}^4x_a(k)
	\end{aligned}
\end{equation*}
where $
\mathcal{A}^1 = \left[ 
\begin{array}{cc}  
	A+\chi_3 BK&  \overline{\beta} BK\\  %第一行元素
	\chi \textbf{I}& \overline{\beta} \textbf{I} %第二行元素
\end{array}
\right],~
\mathcal{A}^2 = \left[ 
\begin{array}{cc}  
	\textbf{0}& BK\\  %第一行元素
	\textbf{0}& \textbf{I} %第二行元素
\end{array}
\right]$, $
\mathcal{A}^3 = \left[ 
\begin{array}{cc}  
	BK & \textbf{0} \\  %第一行元素
	\textbf{I} & \textbf{0} %第二行元素
\end{array}
\right],~
\mathcal{A}^4= \left[ 
\begin{array}{c}  
	BK \\  %第一行元素
	\textbf{I}%第二行元素
\end{array}
\right]$.
Then the security issue under \textcolor{black}{deception attack and DoS attack} for non-switched system can be easily elaborated. 

\begin{corollary}\label{cor1}
	Given scalars $0<  \rho_s <1 $, $\overline{\alpha},~\overline{\beta} \in [0,1]$ and the controller gain  $K$, if there exist a matrix $P>0$ and a positive constant $\epsilon$ such that	
$${\Pi}= \left[ \begin{array}{cc}  
			{\Pi}^{11}& {\Pi}^{12}\\
			*& {\Pi}^{22}
		\end{array}
		\right] \preceq 0$$
	where $\widetilde{\alpha} = ~\overline{\alpha}\left(1-\overline{\alpha}\right),~
	\widetilde{\beta} =~ \overline{\beta}\left(1-\overline{\beta}\right),$
	\begin{align*}
		\Pi^{11}	=~& (\mathcal{A}^1)^TP \mathcal{A}^1+\widetilde{\beta}(\mathcal{A}^2)^TP\mathcal{A}^2  - P + \rho_s P\\&- 2\widetilde{\beta}\left(1-\overline{\alpha}\right)(\mathcal{A}^2)^TP\mathcal{A}^3\\
		&+\left(\widetilde{\alpha}\left(1-\overline{\beta}\right)^2+\widetilde{\beta}\left(1-\overline{\alpha}\right)^2\right)(\mathcal{A}^3)^TP \mathcal{A}^3,\\
		\Pi^{12}=~&-\overline{\alpha}\widetilde{\beta} (\mathcal{A}^2)^TP \mathcal{A}^4+
		\overline{\alpha}\left(1-\overline{\beta}\right)(\mathcal{A}^1)^TP \mathcal{A}^4\\&
		-\widetilde{\alpha}\left(1-\overline{\beta}\right)^2(\mathcal{A}^3)^TP \mathcal{A}^4,\\
		\Pi^{22}	=~&	\overline{\alpha}\left(1-\overline{\beta}\right)(\mathcal{A}^4)^TP \mathcal{A}^4 -\epsilon \textbf{I}.
	\end{align*}
	Then the considered non-switched system is \textcolor{black}{practically stable} in the mean square sense with $\ell$-security level, and
	{\color{black}$	\ell =\max \left\{  {\frac{\tilde{x}^T(0) P\tilde x(0)}{ \lambda_{\min}(P)},~\frac{\epsilon \overline{\gamma}^2}  {\lambda_{\min}(P)\rho_s}} \right\}.$}
\end{corollary}

\begin{remark}
	The criterion in Corollary \ref{cor1} is similar to that revealed in \cite{IET2016}. The main difficulty to extend the results from \cite{IET2016} is the switching signal design, since the attack not only alters the system state, but also has influence on  the consistency of the subsystem and the controller. Moreover, the jump at the switching instant is another challenge in the analysis.
\end{remark}

When there is only DoS attack, the stability criterion degenerates to the scenario  with $\overline{\alpha}=0,~\overline{\gamma}=0$.
\begin{corollary}\label{cor2}
	Given scalars $\mu > 1$, $0<  \rho_s <1 $, $\rho_u > 0$, $\overline{\beta} \in [0,1]$, controller gain $K_p$ and $c= \overline{\beta}\frac{1+\rho_u}{1-\rho_s} < 1$, \textcolor{black}{if there exist  matrices $P_p\succ0,~p\in \mathcal{M}$} such that 
	\begin{align}
		{\Pi}_p&\preceq 0 ,\\
		{\Omega}_{pq}&\preceq 0, \\
		{\Psi}_p&\preceq 0, \\
		P_p &\preceq \mu P_q,~~~~\textcolor{black}{\forall p,q \in \mathcal{M},~p\neq q} ,
	\end{align}
	where $	\widetilde{\beta} =~\overline{\beta}\left(1-\overline{\beta}\right),$
$$\begin{aligned}
		\Pi_p	=~& (\mathcal{A}^1_p)^TP_p \mathcal{A}^1_p+\widetilde{\beta}(\mathcal{A}^2_p)^TP_p\mathcal{A}^2_p  - P_p + \rho_s P_p\\&
		- 2\widetilde{\beta}(\mathcal{A}^2_p)^TP_p\mathcal{A}^3_p
		+\widetilde{\beta}(\mathcal{A}^3_p)^TP_p \mathcal{A}^3_p,\\
		{\Omega}_{pq}=~&(\overline{\mathcal{A}}^1_{pq})^T P_q \overline{\mathcal{A}}^1_{pq} -  P_q - \rho_u P_q,\\
		{\Psi}_p =~& (\widetilde{\mathcal{A}}^{1}_p)^T P_p \widetilde{\mathcal{A}}^{1}_p  -P_p+\rho_s P_p,
\end{aligned}$$
	and the switching signal obeys the rule $\tau_d \geq \tau_d^*=- \frac{\ln\overline{\mu}}{\ln\overline{\rho}_s}$ 
	with $\overline{\mu} = \mu \frac{2-\overline{\beta}-c}{1-c}$. Then the investigated switched system subject to DoS attack is asymptotically  stable in the mean square sense.
\end{corollary}

It is noted that the system can be asymptotically stabilized in the absence of deception attack. By the similar derivation, we can get the results for a switched system under single deception attack, that is, $\overline{\beta}=0$.

\begin{corollary}\label{cor3}
	Given scalars $\mu > 1$, $0<  \rho_s <1 $, $\overline{\alpha}\in [0,1]$, and controller gain $K_p$, \textcolor{black}{if there exist matrices $P_p\succ0,~p\in \mathcal{M}$} and a constant $\epsilon>0$ such that	
	\begin{align*}
		{\Pi}_p= \left[ 
		\begin{array}{cc}  
			{\Pi}_p^{11}& {\Pi}_p^{12}\\
			*& {\Pi}_p^{22}
		\end{array}
		\right] &\preceq 0, \\
		P_p &\preceq \mu P_q,~~~~\textcolor{black}{\forall p,q \in \mathcal{M},~p\neq q},
	\end{align*}
	where $	\widetilde{\alpha} =~\overline{\alpha}\left(1-\overline{\alpha}\right)$,~
	\begin{align*}
		\Pi_p^{11}	=~& (\mathcal{A}^1_p)^TP_p \mathcal{A}^1_p - P_p + \rho_s P_p+\widetilde{\alpha}(\mathcal{A}^3_p)^TP_p \mathcal{A}^3_p,\\
		\Pi_p^{12}=~&\overline{\alpha}(\mathcal{A}^1_p)^TP_p \mathcal{A}^4_p
		-\widetilde{\alpha}(\mathcal{A}^3_p)^TP_p \mathcal{A}^4_p,\\
		\Pi_p^{22}	=~&	\overline{\alpha}(\mathcal{A}^4_p)^TP_p \mathcal{A}^4_p -\epsilon \textbf{I},
	\end{align*}
	and the switching signal satisfies $\tau_d \geq \tau_d^*=- \frac{\ln{\mu}}{\ln\left(1-\rho_s\right)}$. 
	Then the switched system subject to single deception attack is \textcolor{black}{practically stable} in the mean square sense with $\ell$-security index and \\
	\noindent$\ell = {\mu} \max \left\{\textcolor{black}{\frac{\tilde{x}^T(0) P _{\sigma(0)} \tilde x(0)}{ \min_{p\in\mathcal{M}}\{\lambda_{\min}\left(P_p\right)\}},~\frac{\epsilon \overline{\gamma}^2}  {\min_{p\in\mathcal{M}}\{\lambda_{\min}\left(P_p\right)\}\rho_s}} \right\}$.
\end{corollary}

\begin{remark}
	Comparing Corollary \ref{cor2} with Corollary \ref{cor3}, it is obvious that the system under deception attack will converge to  a neighborhood of the origin. In such a case, there is no direct influence on the switching signal, and the switching signal is  similar to that in \cite{DT3}. Furthermore, the switched system with DoS attack will lead to asynchronous behavior between the system mode and the controller mode, which imposes stronger constraint on the switching signal. 
\end{remark}

In what follows, we formulate an explicit expression of the controller gain based on Theorem \ref{thm1}.

\begin{theorem}\label{thm2}
	Given scalars $\mu > 1$, $0<  \rho_s <1 $, $\rho_u > 0$, $\overline{\alpha},~\overline{\beta} \in [0,1]$ and $	c= \overline{\beta}\frac{\overline{\rho}_u}{\overline{\rho}_s} < 1$ with  $\overline{\rho}_s = 1-\rho_s$, $\overline{\rho}_u = 1+\rho_u$,  \textcolor{black}{if there exist matrices $P_p^1\succ0,~P_p^2\succ0$}, $R_p= \left[\begin{array}{c}
		R_p^1\\
		0
	\end{array}
	\right]$, $\Xi_p = \left[\begin{array}{cc}
		\Xi_p^{11}& \Xi_p^{12} \\
0	& \Xi_p^{22}
	\end{array}
	\right]$ and a constant $\epsilon$ such that
	\begin{align}
		{\widetilde{\Pi}}_p= \left[ 
		\begin{array}{cc}  
			{\widetilde{\Pi}}_p^{11}& {\widetilde{\Pi}}_p^{12}\\
			*& {\widetilde{\Pi}}_p^{22}
		\end{array}
		\right] &\preceq 0 , \label{conn1}\\
		\left[
		\begin{array}{cccc}
			-\widetilde{\Omega}_{pq}&0 & \Xi_q E_pA_q &R_q\\
			* & -P_p^2 & 0 & P_p^2\\
			* &* & -\overline{\rho}_u{P^1_p} &0\\
			* & * & * &-\overline{\rho}_u P_p^2
		\end{array}
		\right] &\preceq 0,     \label{conn2}		\\
		\widetilde{\Phi}_p = \left[
		\begin{array}{cc}
			\widetilde{\Phi}_p^{11} & \widetilde{\Phi}_p^{12}\\
			* & \widetilde{\Phi}_p^{22}
		\end{array}
		\right] &\preceq 0     ,\label{conn3}\\
		\text{diag} \{ P_p^1,P_p^2 \} \preceq \mu  ~\text{diag} \{ P_q^1,P_q^2 \}, \forall p,q \in &\mathcal{M},~p\neq q,
	\end{align}
	where  $\Upsilon_{p}^1=\Xi_p E_p A_p +\chi_3 R_p$, $\vartheta_1=	\overline{\alpha}(1-\overline{\beta}) $, $\vartheta_2=\sqrt{\widetilde{\beta}(2-2\overline{\alpha})}$, $\vartheta_3 =\sqrt{\widetilde{\alpha}}(1-\overline{\beta})$, $\vartheta_4= \sqrt{ \widetilde{\beta}(1-\overline{\alpha})(2-\overline{\alpha})} $, $	\widetilde{\Omega}_{pq}=\Xi_q E_p+ (\Xi_q E_p )^T-P_p^1$, 	$E_p = \left[ 
	\begin{array}{cc}
		B_p(	(B^T_pB_p)^{-1} )^T& B^\perp
	\end{array}
	\right]^T  $ with $B^\perp$ being an orthogonal basis of the null space for $B^T$, $	\Lambda_p = {\rm diag} \{\Xi_p E_p + (\Xi_p E_p)^T - P_p^1 , P_p^2\},$
	\begin{align*}
		{\widetilde{\Pi}}_p^{11}& ={\rm{diag}} \left\{-\Lambda_p,-\Lambda_p,-\Lambda_p,-\Lambda_p,-\Lambda_p\right\},\\
		{\widetilde{\Pi}}_p^{12} &= \left[\begin{array}{ccc}
%			\Upsilon_p^1&\Upsilon_p^2&\Upsilon_p^3\\
		\Upsilon_{p}^1 & \overline{\beta} R_p& \vartheta_1 R_p\\
			\chi_3 P_p^2 & 	 \overline{\beta}P_p^2& \vartheta_1P_p^2\\
			0& 	\sqrt{\overline{\alpha} \widetilde{\beta}}  R_p &-	\sqrt{\overline{\alpha} \widetilde{\beta}}  R_p \\
			0& 	\sqrt{\overline{\alpha} \widetilde{\beta}}  P_p^2&-	\sqrt{\overline{\alpha} \widetilde{\beta}}  P_p^2\\
			0& \vartheta_2 R_p  &0\\
			0&\vartheta_2 P_P^2 &0\\
			\vartheta_3 R_p &0&-\vartheta_3 R_p\\
				\vartheta_3 P_p^2 &0&-\vartheta_3 P_p^2\\
		\vartheta_4 R_p &0&0\\
	\vartheta_4 P_p^2 &0&0\\
		\end{array}
		\right], \\
		{\widetilde{\Pi}}_p^{22}& = {\rm{diag}}  \{-(1-\rho_s)P_p, -\epsilon \textbf{I}\},\\
		{\widetilde{\Phi}}_p^{11}& ={\rm{diag}} \left\{-\Lambda_p,-\Lambda_p\right\},\\
		\widetilde{\Phi}_p^{12}& = \left[\begin{array}{ccc}
			\Xi_p E_p A_p + (1-\overline{\alpha}) R_p & \textbf{0} & \overline{\alpha} R_p\\
			(1-\overline{\alpha}) P^2_p & \textbf{0} & \overline{\alpha} P^2_p\\
			\sqrt{\widetilde{\alpha}} R_p & \textbf{0} & \sqrt{\widetilde{\alpha}} R_p\\
			\sqrt{\widetilde{\alpha}} P^2_p & \textbf{0} & \sqrt{\widetilde{\alpha}} P^2_p
		\end{array}
		\right],\\
		{\widetilde{\Phi}}_p^{22} &= {\rm{diag}}  \{-(1-\rho_s)P_p, -\epsilon \textbf{I}\}.
	\end{align*}
	The switched system under \textcolor{black}{deception attack and DoS attack} is \textcolor{black}{practically stable} in the mean square sense with $\ell$-security level.
	Then the controller gain can be calculated by  $		K_p =( \Xi^{11}_p )^{-1} R_p$
	with switching signal satisfying (\ref{taud}). 
	
\end{theorem}
\begin{pf}
	Please see Appendix \ref{app2}. \qed
\end{pf}

\subsection{Mixed-switching control strategy}
In Subsection \ref{subsec1}, it is shown that the system is \textcolor{black}{practically stable} in the mean square sense with $\ell$-security level and $\ell$ is related to the maximum value of deception attack $\overline{\gamma}$. The system converges into a   neighborhood of equilibrium since the value of deception attack $x_a(k)$ may have a noticeable impact on the state by generating an error control signal when the actual state is sufficiently small. 
Fortunately, a remarkable advantage of switched system is that the closed-loop system can achieve stability even if all subsystems are unstable. However, this method does not perform well when the state is far from the equilibrium. 
In this part, we give a novel control strategy to enhance the resilience of switched system by incorporating the state-switching design method. 

{\color{black}
Due to  the randomness of attacks, the switching condition for state switching and time switching will be triggered frequently if it only relates to the state information, leading to undesired performance such as oscillations. 
How to reduce the number of switchings between state switching and time switching and achieve asymptotic stability is challenging. }

First, we introduce two strategies consisting of controller and switching signal.

\textbf{\textit{Strategy 1: }} The controller is 
\begin{equation}\label{cont2_1}
	u(k) = K_{\sigma(k)} \hat{x}(k) 
\end{equation}
where 
$\hat x(k) = (1-\alpha(k))(1-\beta(k))x(k) + \alpha(k)(1-\beta(k)) x_a(k)$
and the next switching instant satisfies
\begin{equation}\label{switching_law}
	k_{s+1}	\geq k_s+ \tau_{d1} 
\end{equation}
with $\tau_{d1}$ being given in Theorem \ref{thm3}.

\textbf{\textit{Strategy 2: }} The controller is 
\begin{equation}\label{cont2_2}
	u(k) = 0
\end{equation}
and the next switching instant is 
\begin{equation}\label{switching_law2}
k_{s+1}		\triangleq \inf \{k \geq k_s+\tau_{d2}:~\varphi(k) > 0 \} 
\end{equation}
where  $\varphi(k) = 	x^T(k) Q_p x(k)
- \min \limits_{q \in \mathcal{M}, q\neq p} \mu x^T(k) Q_q x(k) 
$ ($p,q \in \mathcal{M}, p\neq q$) and $Q_p$ for all $p\in \mathcal{M}$ is a  positive define matrix to be designed.

Our control strategy switches to \textbf{\textit{Strategy 1}} once $	\|x(k)\| >  \overline{\gamma}^2$ and the switching conditions from \textbf{\textit{Strategy 1}} to \textbf{\textit{Strategy 2}} are $	\|x(k)\| \leq \overline{\gamma}^2$ and $k	\geq k_s+ \tau_{d1} $. Notably, without the second condition, the control method will switch back and forth between the two ones, which may deteriorate control performance and is not expected to happen in practice.

{\color{black}\begin{remark}
	The connection between the switching condition for two switching laws and the deception attack is inherent in the nature of the system. The state switching law demonstrates optimal performance when the state remains within the bounds of the deception attack. Additionally, to minimize frequent switches, we advocate the imposition of an additional time constraint on the switching signal when the two laws coexist. The adoption of a time switching constraint proves practical for limiting the switching frequency, ensuring ease of implementation in real-world scenarios.
	\end{remark}}

Accordingly,  the closed-loop system under control \textbf{\textit{strategy 1}}   has the form
\begin{align}
	&	x(k+1)\label{clo_1}  \\
		=~ & {\mathcal{A}}_{\sigma(k)} x(k) + (1-\alpha(k))(1-\beta(k)) B_{\sigma(k)}K_{\sigma(k)}x(k)\notag
		\\&+  \alpha(k)(1-\beta(k))  B_{\sigma(k)}K_{\sigma(k)}x_a(k) \notag
	\end{align}
and  the closed-loop system with control \textbf{\textit{strategy 2}} is 
\begin{equation}\label{clo_3}
	x(k+1)  = A_{\sigma(k)} x(k).
\end{equation}

\begin{remark}
	Different from the control signal  in section \ref{subsec1}, the control signal turns to be zero rather than the value at the latest successfully transmitted instant when the DoS attack occurs. Consequently,  the mismatching behavior will not happen.
\end{remark}

In the sequel, we focus on the design of switching signal in terms of the information of time and state for controller (\ref{cont2_1}). The dwell time constraint  leads to a slow frequency, which reduces the number of asynchronous behavior and the high frequency switching. It is noted that the switching law has a mixed form, which combines the dwell time approach as that in Section \ref{subsec1} and the state-dependent switching rule in  \cite{Lu2016}.  The aim of $\tau_{d1}$  is to make sure that the decrease rate of stable subsystem can compensate the energy increase  at the switching instants. $\tau_{d2}$ guarantees that the switching frequency has an upper bound and $\varphi(k) \leq 0$ implies that the energy decreases at the switching instants.

\begin{theorem}\label{thm3}
	Given scalars $0<\rho_s < 1$, $\rho_u>0$, $\mu\geq 1$, $\mu_1\geq 1$ and $\mu_2\geq 1$ and matrices $K_p$. If there exist matrices ${P}_p\succ0 $ and $Q_p$, $p\in \mathcal{M}$ and a positive constant $\epsilon$ such that 
	\begin{align}
{\varLambda}_p= \left[ 
\begin{array}{cc}  
\varLambda_p^{11}&\varLambda_p^{12}\\
* &\varLambda_p^{22}
\end{array}
\right] \preceq 0 ,\label{thm4_1} \\
	\mu_1 {P}_q  - {P}_p \succeq 0,~ p,q \in \mathcal{M}, p\neq q,    \label{thm4_2}\\
		A^T_pQ_p A_p -Q_p+ \lambda(\mu Q_q - Q_p) \preceq 0,     \label{thm4_3} \\
	\mu Q_{p}  - {P}_q\preceq 0,~ p,q \in \mathcal{M},    \label{thm4_4} \\
 \mu_2 Q_{p}  - {P}_q\succeq 0,~ p,q \in \mathcal{M},   \label{thm4_5} 
	\end{align}
where 
\begin{align*}
		\varLambda_p^{11} 
	=~& A^T_p P_p A_p-P_p + 2(1-\overline{\alpha})(1-\overline{\beta}) A^T_p P_p B_pK_p \\
	&+(1-\overline{\alpha})(1-\overline{\beta})  (B_pK_p)^TP_pB_pK_p+\rho_sP_p,\\
	\varLambda_p^{12} =~& \overline{\alpha}(1-\overline{\beta}){{A}}^T_pP_p B_pK_p,\\
	\varLambda_p^{22} =~&	\overline{\alpha}\left(1-\overline{\beta}\right)(B_pK_p)^TP_p B_pK_p-\epsilon \textbf{I},
\end{align*}
and the switching signal satisfies\\
\noindent $\tau_{d1} \geq \tau_{d1}^* =\max\left\{-\frac{\ln \mu_1}{\ln (1-\rho_s)},-\frac{\ln \mu_2}{\ln (1-\rho_s)}\right\}$
 and 
 \begin{equation}\label{state_switching2}
 	\tau_{d2} \leq \tau_{d2}^* =\frac{\ln \mu}{\ln (1+\lambda)}.
 \end{equation}
Then switched system under \textcolor{black}{deception attack and DoS attack} with mixed-switching strategy is globally asymptotically stable in mean square sense.
\end{theorem}

\begin{pf}
	Please see Appendix \ref{app3}.
\end{pf}

{\color{black}
	\begin{remark}
	In this part, we devise a novel control strategy which combines the feedback control and the switching control by designing appropriate switching signal. Similar to the meaning of Theorem 1, LMIs (\ref{thm4_1}) and (\ref{thm4_2}) are the constraints for Strategy 1. The first one makes sure that the Lyapunov function for the system under attacks decreases, and the second one guarantees that the energy at  switching instant has a bounded growth  when time-dependent switching signal is adopted. The state-dependent switching signal is constrained by LMI (\ref{thm4_3}). Finally, LMIs (\ref{thm4_4}) and (\ref{thm4_5}) ensure the jump at strategy switch instants being bounded.
\end{remark}}

{\color{black}
\begin{remark}
	The switched system under time-dependent switching law and feedback controller is practically stable due to the deception attack. More specifically, the influence of deception attack will be dominated near the equilibrium. In other words, the controller signal cannot stabilize the system when  controller signal has similar amplitude to deception attack signal. Therefore, to reduce the influence of deception attack, we resort to the switching signal to stabilize the system once the deception attack clearly affects the system.
\end{remark}}
{\color{black}	\begin{remark}
			The theoretical merits of the mixed switching control strategy are outlined as follows: 1) {Global Asymptotic Stability}: In contrast to Theorem 2, which utilizes a time-dependent switching law, the system exhibits global asymptotic stability to the equilibrium even in the presence of a deception attack—an occurrence that is unattainable in a non-switched system under a peak-bounded deception attack. 2) {Convergence Speed}: Compared with the state-dependent switching law, the state converges rapidly with fewer instances of switching. 3)	{Optimal Utilization of Switching Laws}: The mixed switching strategy capitalizes on the advantages of both time-dependent and state-dependent switching laws. Time-dependent switching employs feedback control to stabilize the system, while state-dependent switching excels in stabilizing the closed-loop system through a well-designed switching signal, even when individual subsystems are unstable. However, the latter performs less effectively when the state is far from the equilibrium.
	\end{remark}
 }

{\color{black}
	\begin{remark}
		Two switching laws to stabilize switched system have been presented. The first one is time-dependent switching law, which is easy to utilize since only the controller needs to be designed. Note that it is nontrivial to extend the results from  non-switched system to switching system by considering both deception attack and DoS attack. That is because the switching signal constraint should be investigated in connection with the attack parameters and the asynchronous behavior which is related to the probability of DoS attack. However,  the controller has no resilience to  deception attack. 
		Another one is mixed-switching law, which can guarantee the asymptotic stability of the switched system under deception attack. To the best of the authors' knowledge, it is the first attempt to enhance resilience against deception attack with the help of switching law. 
	\end{remark}
}

\section{Simulations}\label{sec_sim}
This section shall carry out three numerical examples to verify the theoretical finding of the proposed setup.

\subsection{Example 1}

Consider a switched system consisting of two subsystems
\begin{equation*}
\begin{aligned}
		A_1 = \left[\begin{array}{cc}
		~~0.88&~~0.23\\
		~~0.84&-0.47
	\end{array}\right],~
	B_1 = \left[\begin{array}{cc}
		-0.77&-0.33\\
		-0.31&~~0.50
	\end{array}\right],\\
	A_2 = \left[\begin{array}{cc}
		~~0.99&-0.08\\
		-0.39&-0.33
	\end{array}\right],~
	B_2 = \left[\begin{array}{cc}
		~~0.47&~~0.31 \\
		~~0.60&-0.55
	\end{array}\right],
\end{aligned}
\end{equation*}
with parameters $\rho_s = 0.15,~\rho_u =0.3,~\overline{\gamma} = 0.13, ~\alpha = 0.13, ~\beta = 0.13, ~\mu=1.1$. By Theorem \ref{thm2}, the controller gains can be calculated as
$$
\begin{aligned}
	K_1 = \left[\begin{array}{cc}
		0.7848 &   0.0864\\
		0.2825    &0.3376\\
	\end{array}\right],~
	K_2 = \left[\begin{array}{cc}
		-1.6734  &  1.9332\\
		~~0.0410  &  0.1545\\
	\end{array}\right],
\end{aligned}
$$
with $\epsilon=1.7389$, $\tau_d^*  =    8.1838$. 
We conduct the Monte Carlo experiments with 100 times. Here the state norm  
$\|\chi(s)\| \triangleq \frac{\|\sum_{i = 1}^{100} x^i(s)\|}{100}$,	where $x^i(s)$ is the state at time $s$ in the $i$th run.

The switching signal is depicted in Fig. \ref{Fig_switching}. To show the asynchronous behavior cased by DoS attack, the DoS attack and the controller mode are also plotted  in Fig. \ref{Fig_switching}.
Fig. \ref{Fig_norm} shows the  norm of state and the bound $\psi$. It is clear that the state is bounded within $\psi$ which is indicated by the red line.
Figs. \ref{Fig_alpha}--\ref{Fig_level} portray how parameters affect the state and the asymptotic bound.  It is clear from Fig. \ref{Fig_alpha}, the state jitter is more severe as $\overline{\alpha}$ increases  and the bound becomes larger with the increase of attack frequency. The effect of DoS attack probability on the state and the asymptotic bound is illustrated in Fig. \ref{Fig_beta}. Specially, the controller gain is chosen to make the subsystem stabilizable and no switching occurs during this simulation in the fourth subgraph of Fig. \ref{Fig_beta}. We can see that the state still diverges even the LMIs are not feasible. 
There is a certain conservativeness for the obtained conditions due to inequality scaling in the derivation process. More specific,  it is found that the system can be stabilizable yet the LMIs become infeasible when $\overline{\beta} = 0.4$ and other parameters are chosen unchanged.} Analogously, a larger attack bound $\overline{\gamma}$ leads to an increase of asymptotic bound, which has strong impact than that deduced by attack probability in Fig. \ref{Fig_level}.

\begin{figure}
\begin{minipage}{0.45\linewidth}
\vspace{0.2cm}
	\centering
	\includegraphics[width=1\linewidth]{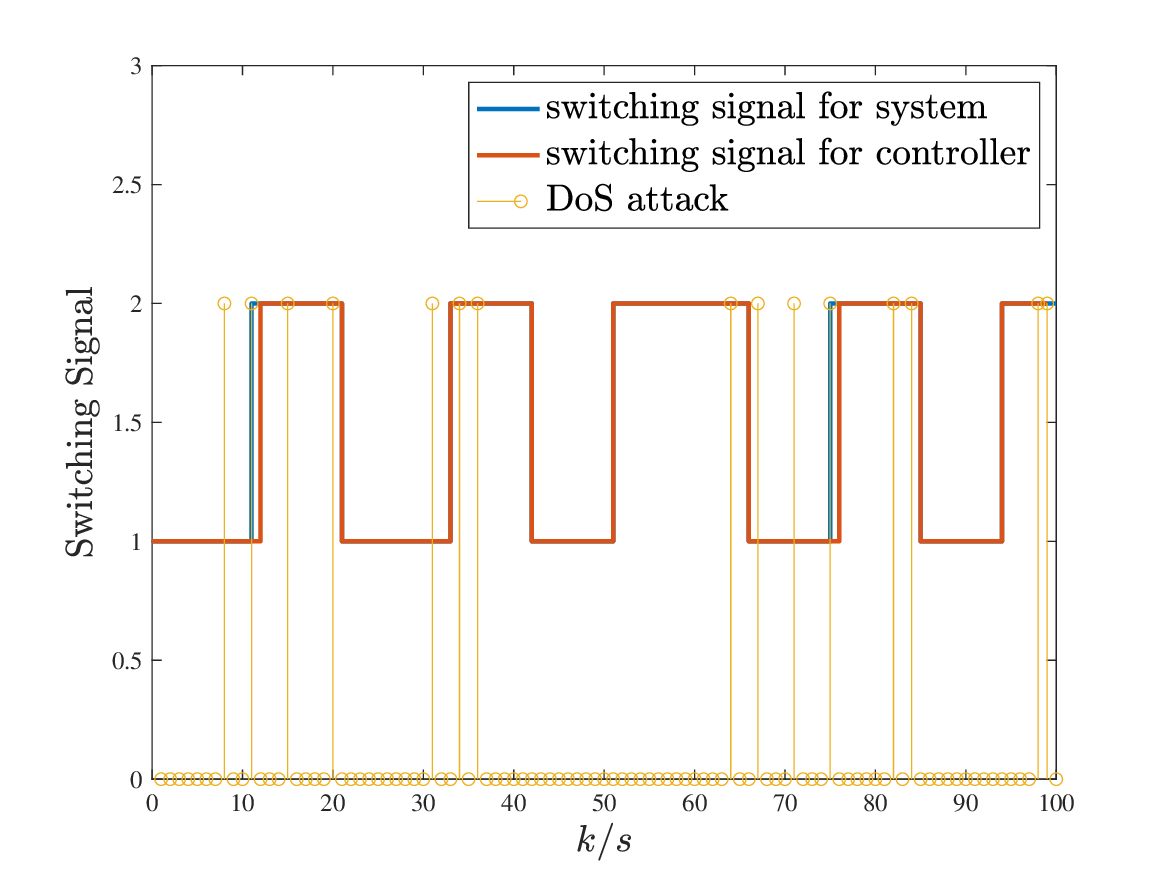}
	\caption{\textcolor{black}{The switching signal for Example 1}}
	\label{Fig_switching}
\end{minipage}~~~
\begin{minipage}{0.47\linewidth}
	\centering
	\includegraphics[width=1\linewidth]{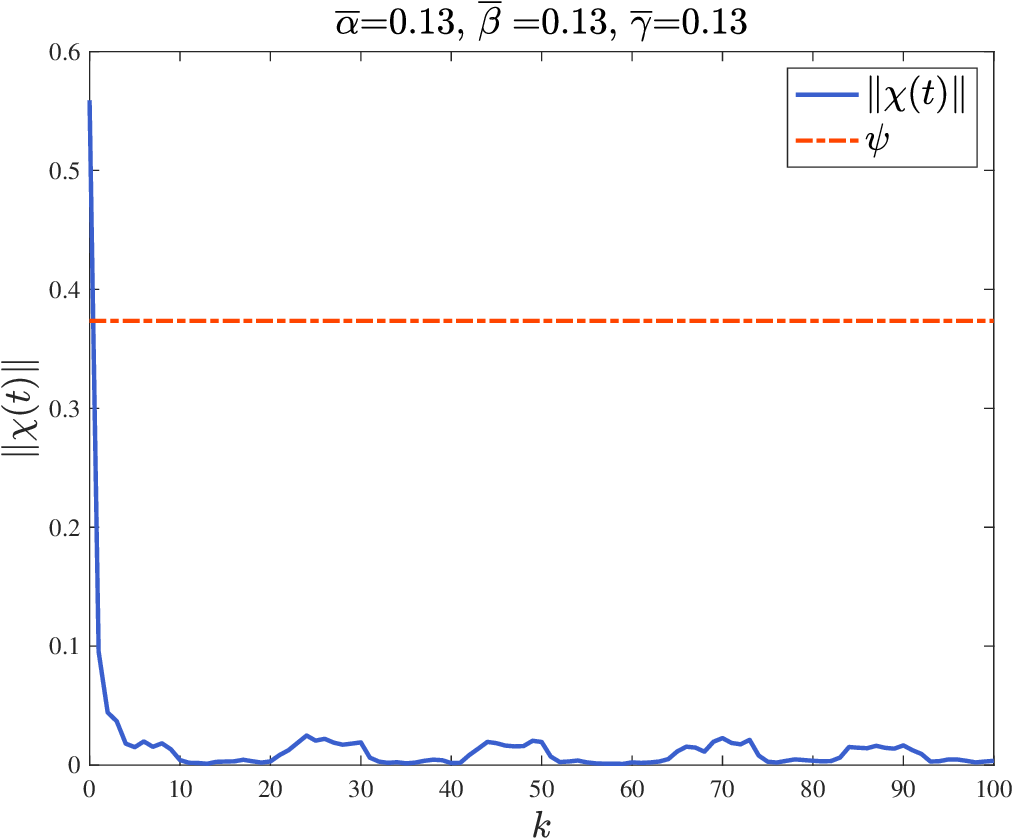}
	\caption{The norm of state and the bound $\psi$ for Example 1}
	\label{Fig_norm}
\end{minipage}
\end{figure}

\begin{figure}[t]
\centering
\subfigure{\includegraphics[width=3.9cm]{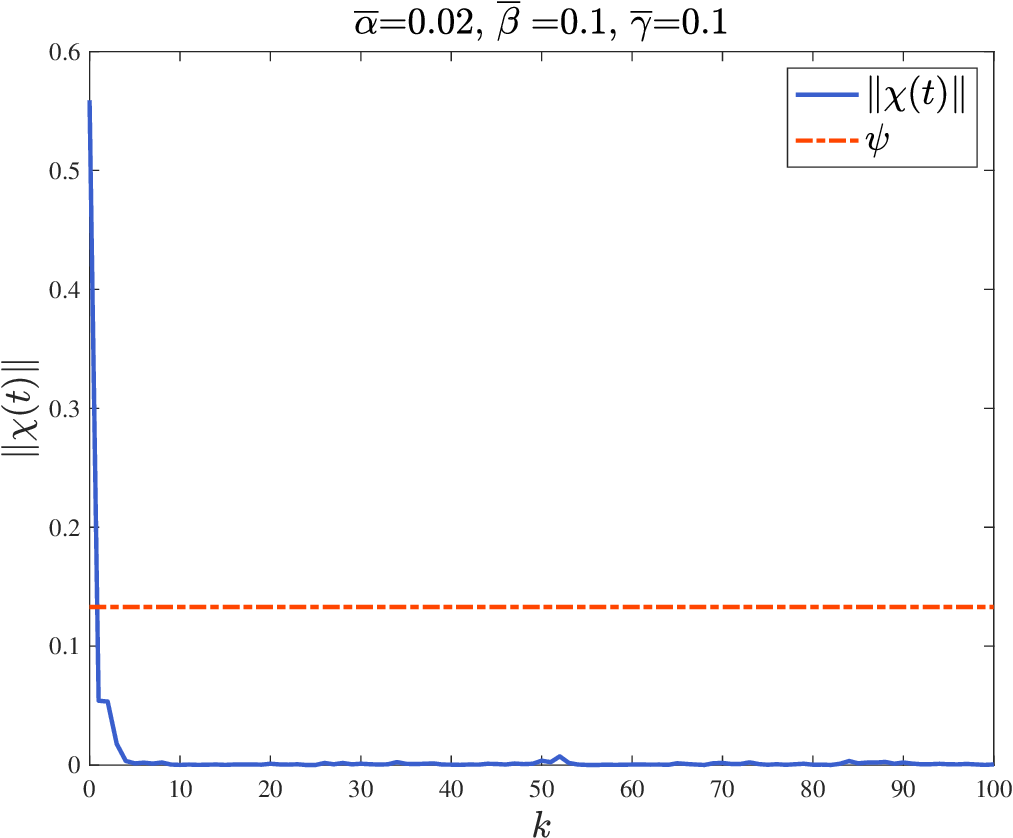}}
\subfigure{\includegraphics[width=3.9cm]{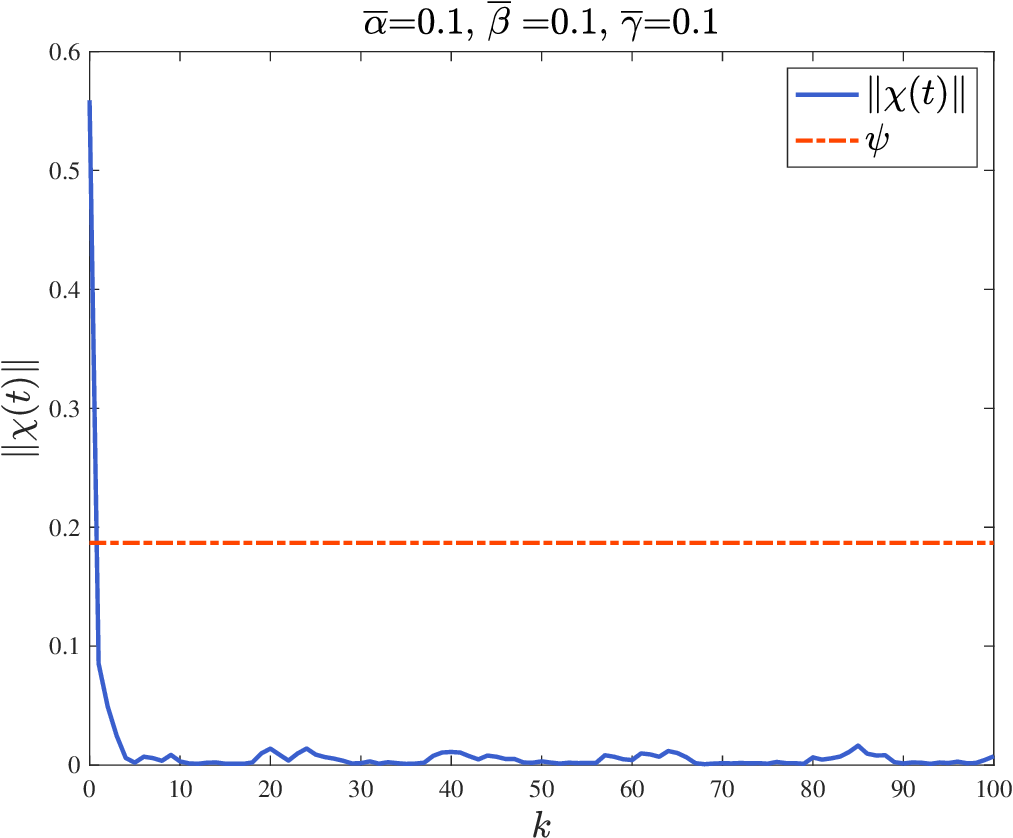}}
\subfigure{\includegraphics[width=3.9cm]{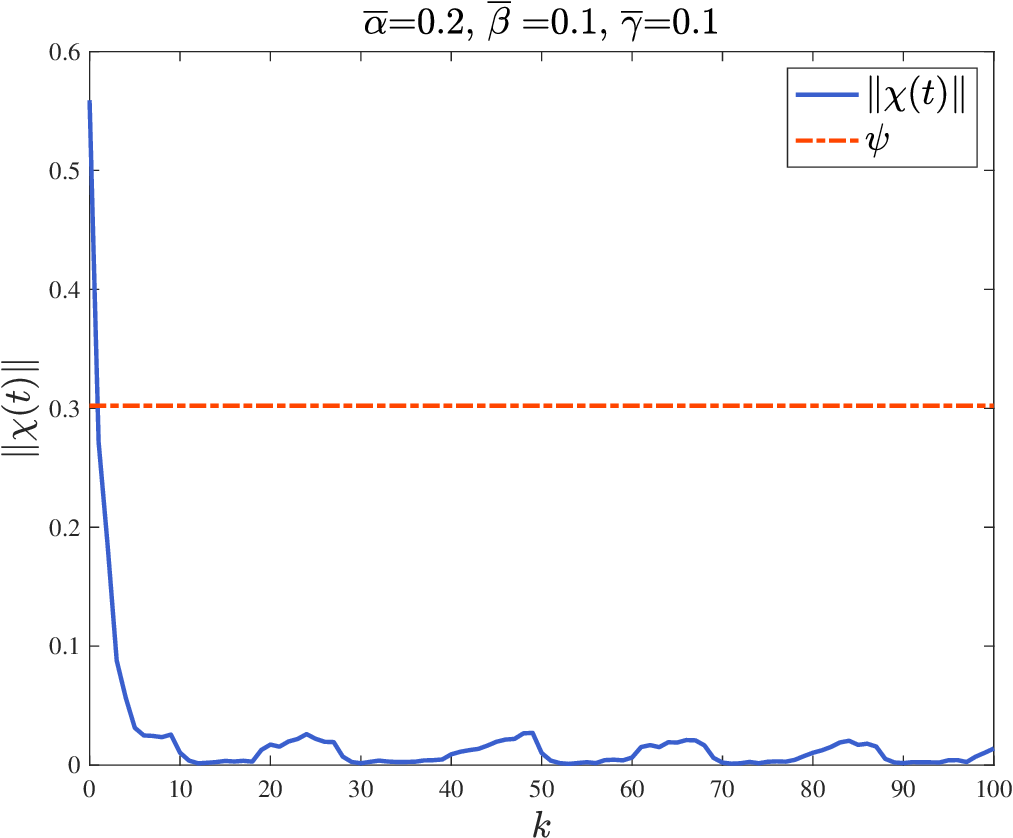}}
\subfigure{\includegraphics[width=3.9cm]{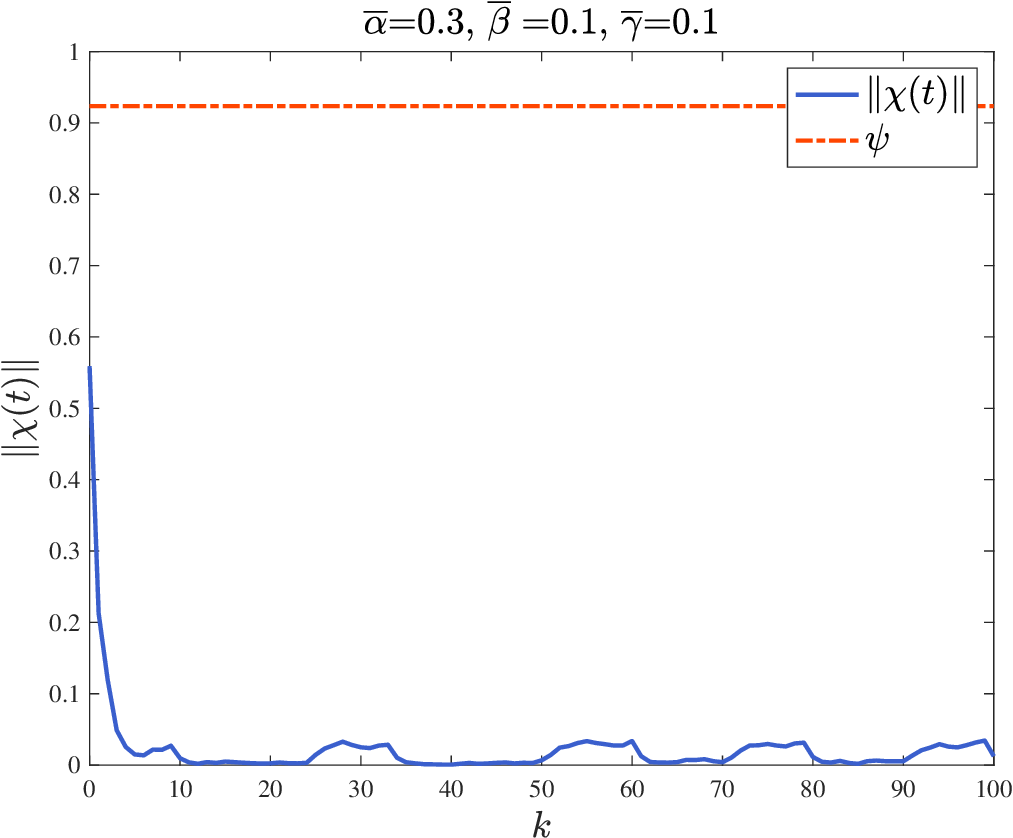}}
\caption{The norm of state and the bound $\psi$ under  different deception attack probabilities with the same DoS attack probability and deception attack level for Example 1.}
\label{Fig_alpha}
\end{figure}

\begin{figure}[t]
\centering
\subfigure{\includegraphics[width=3.9cm]{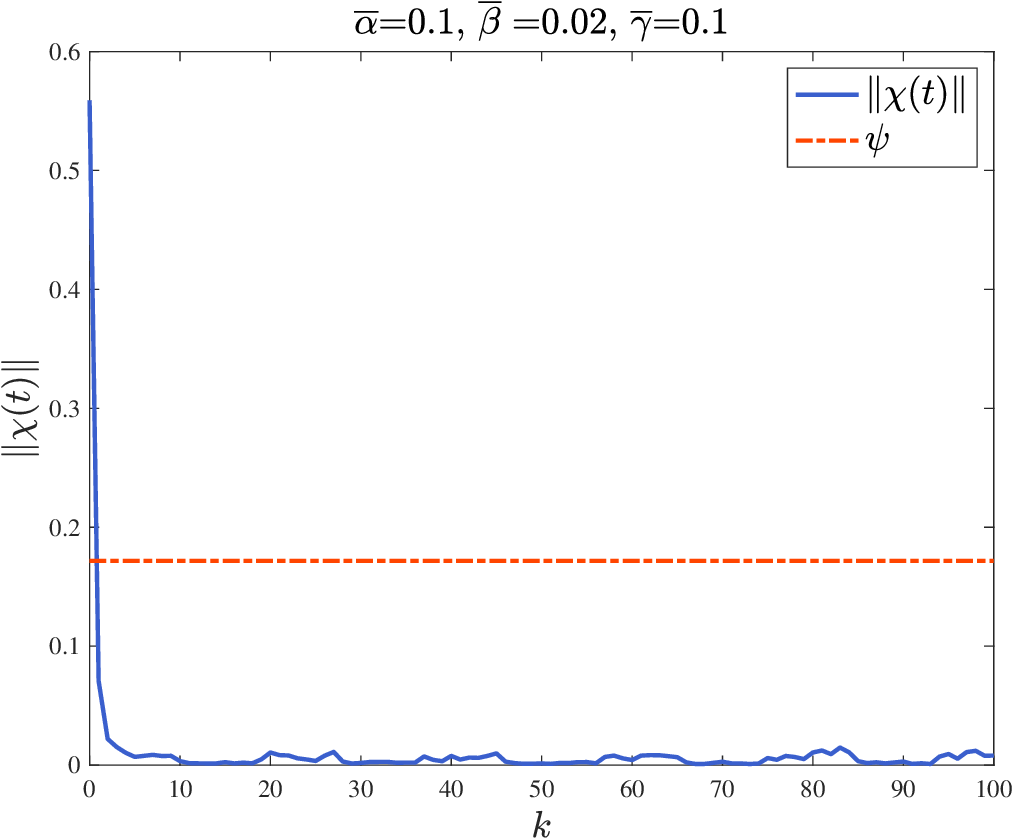}}
\subfigure{\includegraphics[width=3.9cm]{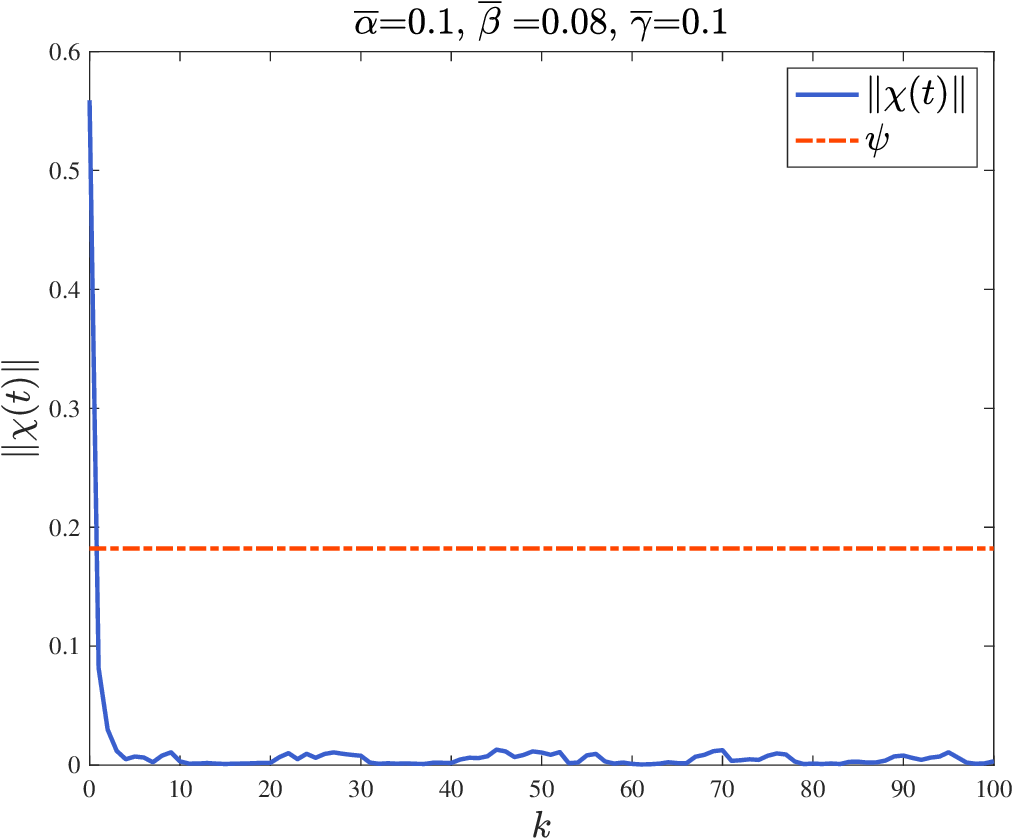}}
\subfigure{\includegraphics[width=3.9cm]{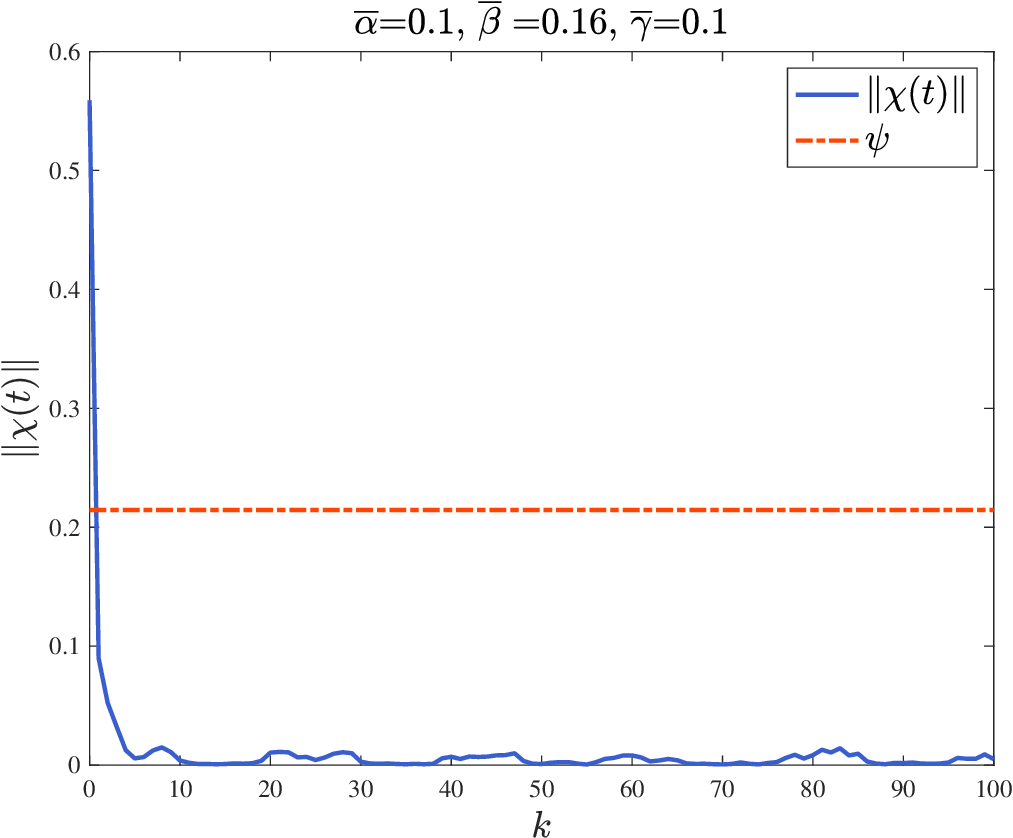}}
\subfigure{\includegraphics[width=3.9cm]{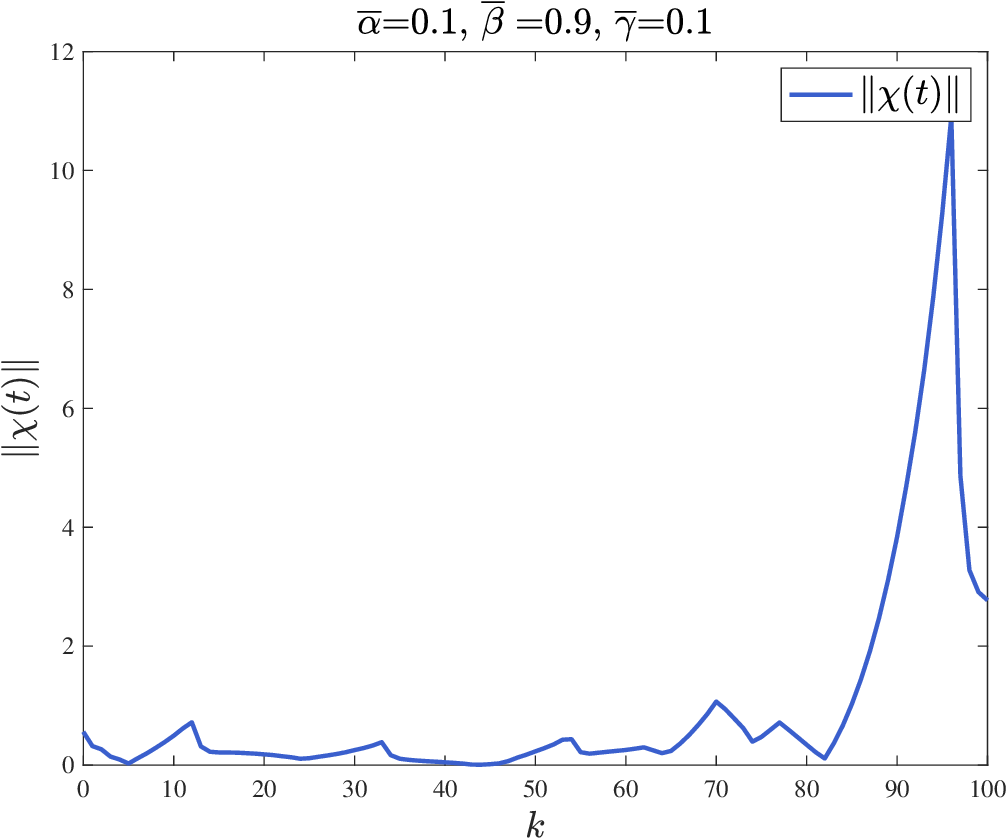}\label{Fig_beta_d}}
\caption{The norm of state and the bound $\psi$ under  different DoS attack probabilities with the same deception attack probability and level for Example 1. }
\label{Fig_beta}
\end{figure}

\begin{figure}[t]
\centering
\subfigure{\includegraphics[width=3.9cm]{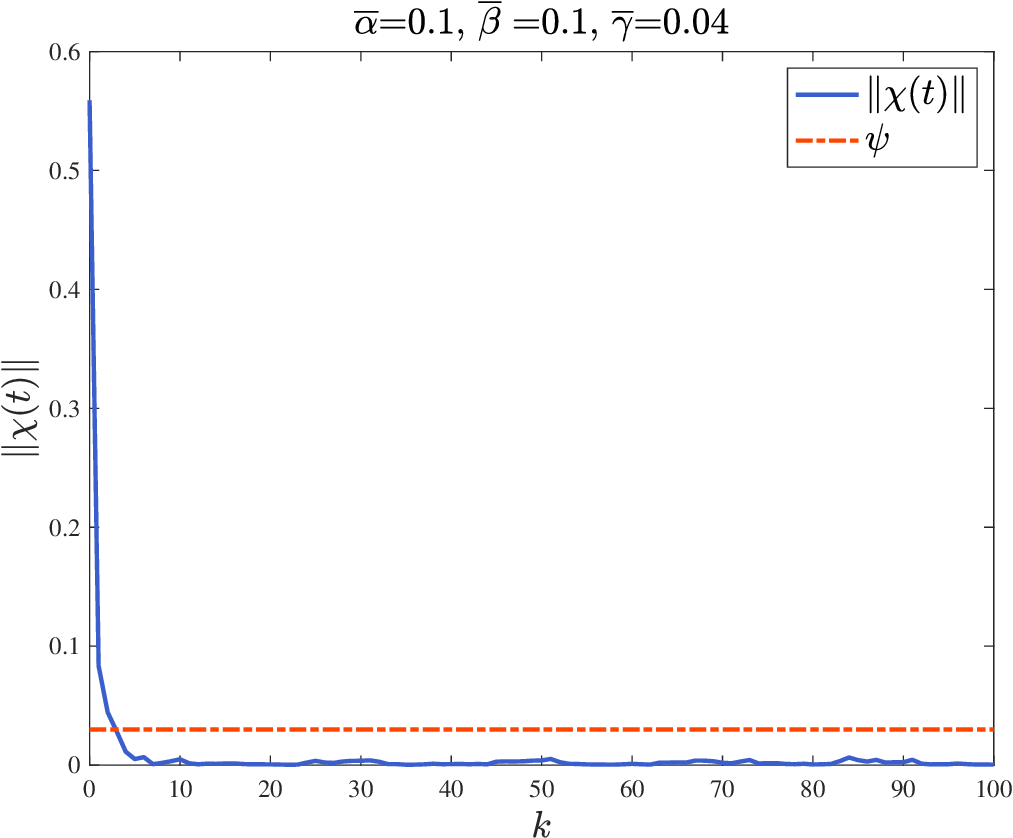}}
\subfigure{\includegraphics[width=3.9cm]{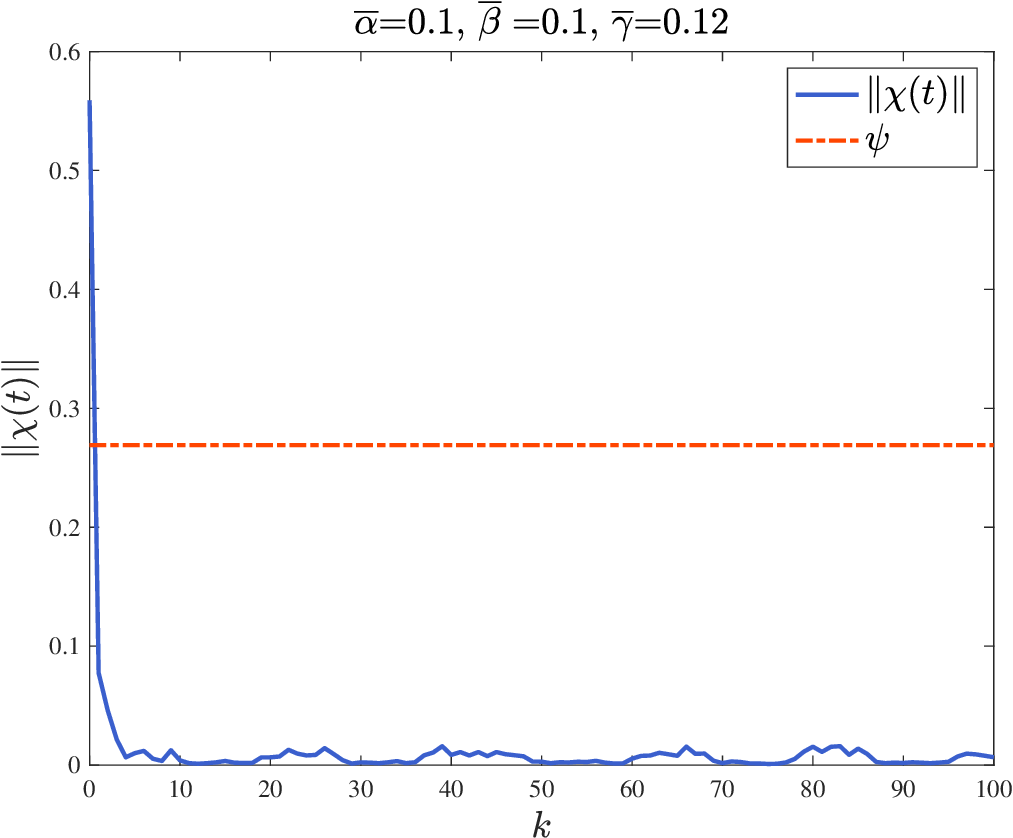}}
\subfigure{\includegraphics[width=3.9cm]{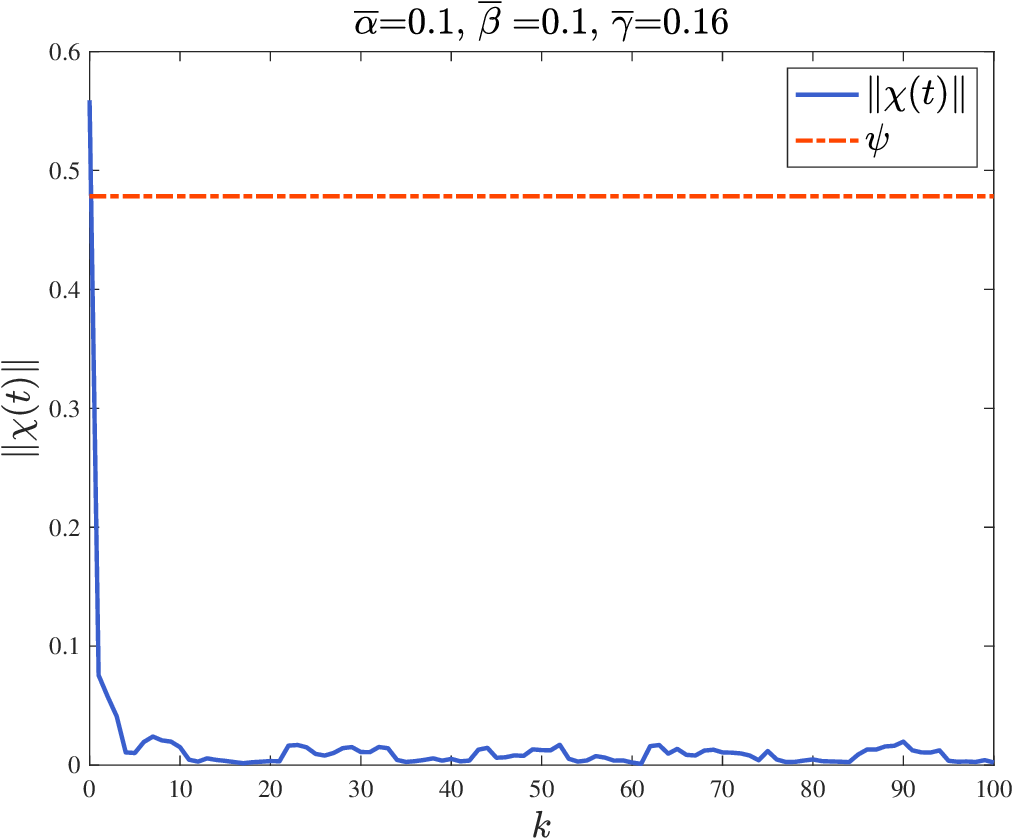}}
\subfigure{\includegraphics[width=3.9cm]{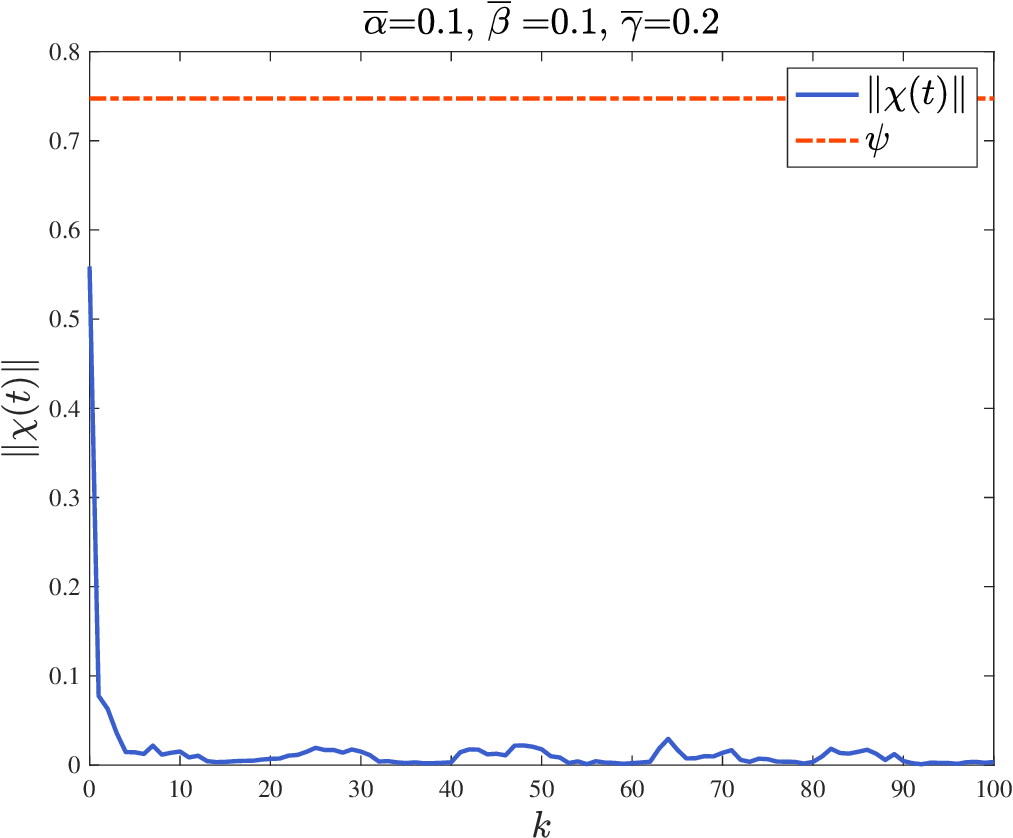}}
\caption{The norm of state and the bound $\psi$ under different deception attack levels with the same probability of DoS attack and deception attack for Example 1.}
\label{Fig_level}
\end{figure}

\begin{table}  
\centering
\scriptsize
{ 	\caption{$\psi$ and $\ell$ with different deception attack probabilities}  
	\label{Tab_alpha}  
	\begin{tabular}{ccccccc}  
		\hline  
		\hline 
		\specialrule{0em}{1pt}{1pt}
		$\overline{\alpha}$&	0&0.02&0.04&0.06&0.08&0.10 \\  
		$\psi$&	0.2262&0.2319&0.2388&0.2471&0.2564&0.2669 \\  
		$\ell$& 0.8506&0.8469&0.8428&0.8383&0.8335&0.8284\\
		\hline
		\specialrule{0em}{1pt}{1pt}
		$\overline{\alpha}$&0.12&0.14&0.16&0.18&0.20&0.22 \\ 
		$\psi$&0.2783&0.2908&	0. 3042& 0.3188&0.3348&0.3526\\
		$\ell$&0.8230&0.8173&	0.8114&0. 8052& 0. 7987&0.7918\\
		\hline 
		\hline 
\end{tabular}}

\end{table}

\begin{table}  
\centering
\scriptsize
{ 	\caption{$\psi$ and $\ell$ with different DoS attack probabilities}  	\label{Tab_beta}
	\begin{tabular}{ccccccc}  
		\hline  
		\hline
		\specialrule{0em}{1pt}{1pt}
		$\overline{\beta}$&	0&0.02&0.04&0.06&0.08&0.10 \\  
		$\psi$&	0.2540&0.2560&0.2582&0.2607&0.2636&0.2669 \\  
		$\ell$& 0.8309&0.8306&0.8303&0.8297&0.8292&0.8284\\
		\hline
		\specialrule{0em}{1pt}{1pt}
		$\overline{\beta}$&0.12&0.14&0.16&0.18&0.20&0.22 \\ 
		$\psi$&0.2708&	0.2754&0.2813&0.2892&0.3008&0.3214\\
		$\ell$&0.8270&0.8249&0.8216&0.8264&0.8084&0.7953\\
		\hline 
		\hline 
\end{tabular}  }
\end{table}

\begin{table}  
\centering
\scriptsize
{ 	\caption{$\psi$ and $\ell$ with different deception attack levels}  	\label{Tab_level}
	\begin{tabular}{ccccccc}  
		\hline  
		\hline  
		\specialrule{0em}{1pt}{1pt}
		$\overline{\gamma}$&0.02&0.04&0.06&0.08&0.10&0.12 \\  
		$\psi$&	0.0107&0.0427&0.0961&0.1708&0.2669&0.3843 \\  
		$\ell$& 0.8284&0.8284&0.8284&0.8284&0.8284&0.8284\\
		\hline
		\specialrule{0em}{1pt}{1pt}
		$\overline{\gamma}$&0.14&0.16&0.18&0.20&0.22 &0.24\\ 
		$\psi$&0.5231&0.6832&	0. 8647& 1.0676&1.2917&1.5373\\
		$\ell$&0.8284&0.8284&	0. 8647&1.0676& 1.2917&1.5373\\
		\hline 
		\hline  
\end{tabular}  }
\end{table}

Tables \ref{Tab_alpha}--\ref{Tab_level} reveal  the relationship between the attack parameters $\overline{\alpha}$, $\overline{\beta}$, $\overline{\gamma}$  and the security level $\ell$, $\psi$.
It is obvious that $\psi$ increases with a larger attack intensity ($\overline{\alpha}$, $\overline{\beta}$, $\overline{\gamma}$). However, the security level $\ell$ is closely related to the initial value of Lyapunov function, which is determined by  matrices $P_p$.  It varies within a small range that is clearly demonstrated in Table \ref{Tab_level}. The above simulations verify the influence of parameters on the security performance, which is consistent with the statement in Remark \ref{rm_canshu}. Moreover, when the initial value of Lyapunov function is less than the bound $\psi$, the security level $\ell$ is equal to $\psi$.

\subsection{Example 2}

Consider the switched system with two subsystems
\begin{equation*}
	A_1 = \left[\begin{array}{cc}
		-0.35&~~0.70\\
		~~0.92&~~0.56
	\end{array}\right],~
	B_1 = \left[\begin{array}{cc}
		~~0.48&~~0.51\\
		-0.79&~~0.06
	\end{array}\right],
\end{equation*}
\begin{equation*}
	A_2 = \left[\begin{array}{cc}
		~~0.96&~~0.33\\
		~~0.36&-0.34
	\end{array}\right],~
	B_2 = \left[\begin{array}{cc}
		-0.50&-0.96 \\
		~~0.72&~~0.51
	\end{array}\right],
\end{equation*}
with parameters $\rho_s = 0.1$, $\lambda =1$, $\mu =1.05$, $\mu_1 = 1.1$ and $\mu_2 = 1.7$. The controller gain are 
$$
\begin{aligned}
	K_1 = \left[\begin{array}{cc}
		~~0.9854&~~0.2560\\
		~~0.5152&-2.4460\\
	\end{array}\right],~
	K_2 = \left[\begin{array}{cc}
		-1.6440&~~1.9572\\
		~~1.6131&-0.8412\\
	\end{array}\right].
\end{aligned}
$$

\begin{figure}
	\begin{minipage}{0.48\linewidth}
		\vspace{0.2cm}
			\centering
		\includegraphics[width=\linewidth]{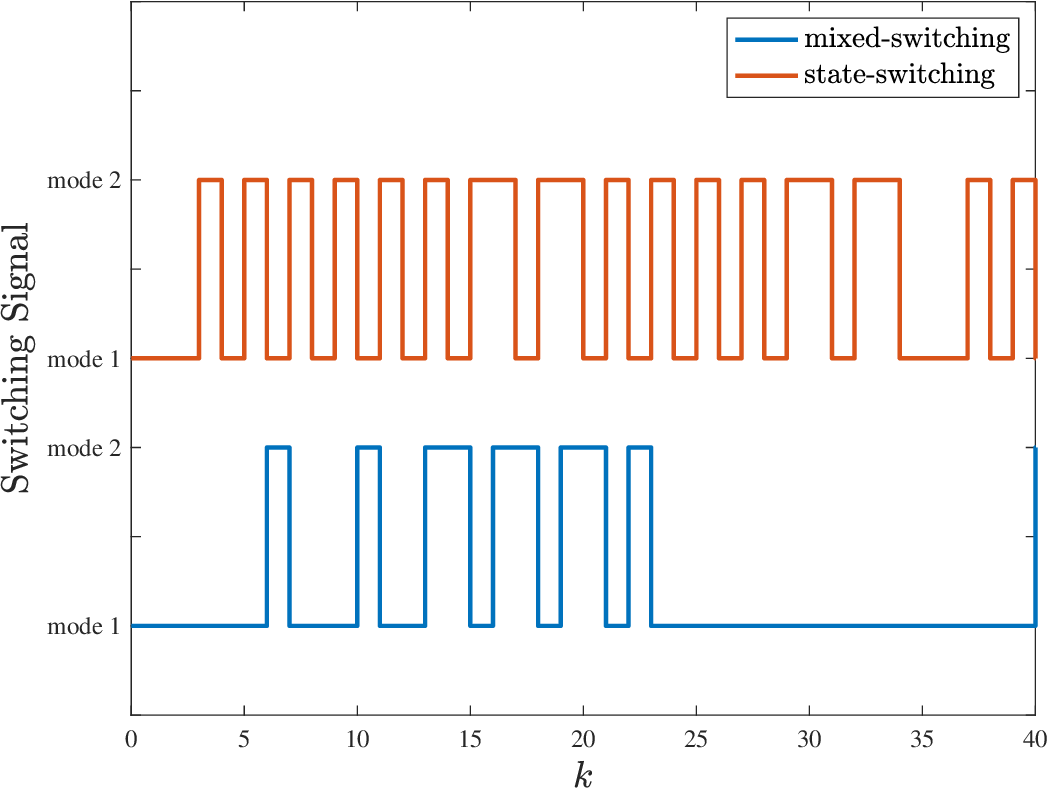}
		\caption{The mixed-switching and state switching signals for Example 2}
		\label{fig:switching}
	\end{minipage}
\begin{minipage}{0.46\linewidth}
	\centering
\includegraphics[width=\linewidth]{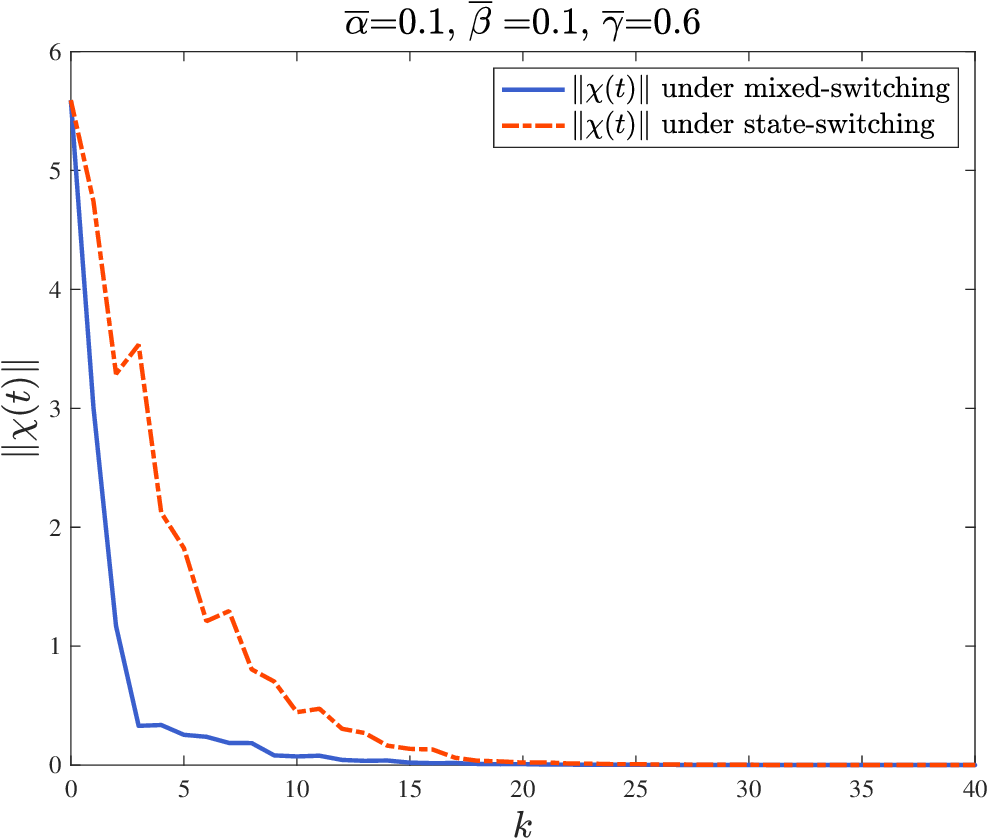}
\caption{The norm of system state under different switching laws for Example 2}
\label{fig:normstate}
\end{minipage}
\end{figure}

\begin{figure}
	\centering
	\includegraphics[width=0.5\linewidth]{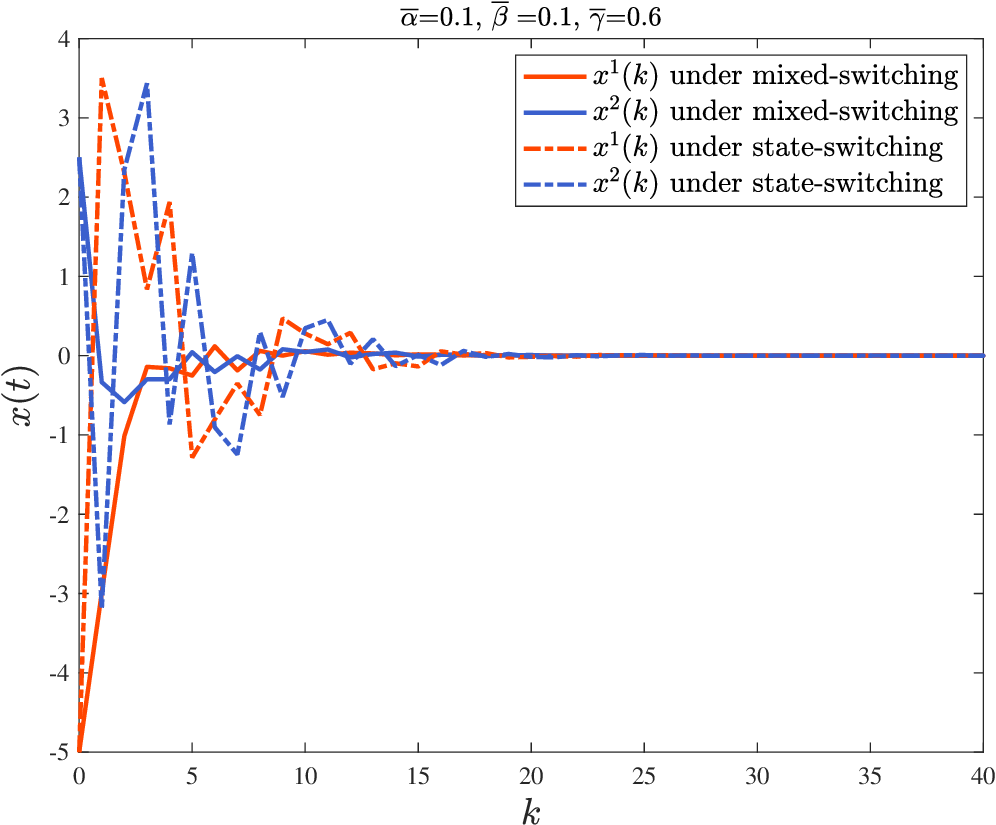}
	\caption{The state trajectories under different switching strategies for Example 2}
	\label{fig:state}
\end{figure}

In this simulation, we compare the mixed-switching  law incorporating both state and time information  with the state-dependent switching law
\begin{equation}\label{state_switching}
k_{s+1}	\triangleq \inf \{k \geq k_s:\varphi(k) \leq 0 \} 
\end{equation}
\textcolor{black}{Fig. \ref{fig:switching} illustrates the switching signals for two cases, demonstrating that the system using a state-switching law has a higher frequency compared to the mixed-switching scheme.  Fig. \ref{fig:normstate} and Fig. \ref{fig:state} depict the state norm and trajectories, indicating superior performance of the mixed-switching control strategy under DoS and deception attacks. Concurrently analyzing Figs. \ref{fig:normstate}-\ref{fig:state}, the state-switching strategy exhibits a faster switch, particularly when the state is far from zero.
}	
	
	{\color{black}
	Figs. \ref{Fig_norm}-\ref{Fig_level} and Fig. \ref{fig:normstate} highlight that the system under the mixed-switching control strategy is asymptotically stable, whereas it achieves practical stability under time-dependent switching in the presence of DoS and deception attacks. This suggests that the mixed-switching control strategy outperforms single time-dependent switching or state-dependent switching.}

\color{black}

\subsection{Example 3}

\begin{figure}
\begin{minipage}{0.45\linewidth}
		\centering
	\includegraphics[width=0.9\linewidth]{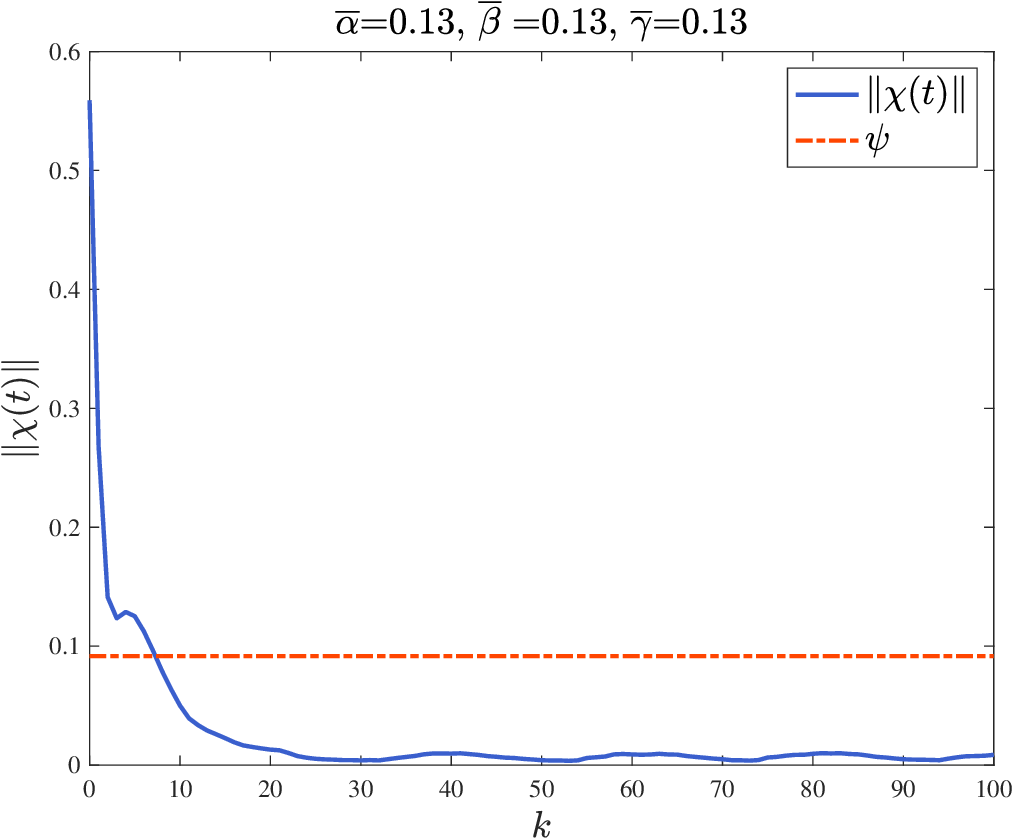}
	\caption{\textcolor{black}{The norm of state and the bound $\psi$ for Example 3}}
	\label{Fig_norm1}
\end{minipage}~~
\begin{minipage}{0.49\linewidth}
	\centering
\includegraphics[width=\linewidth]{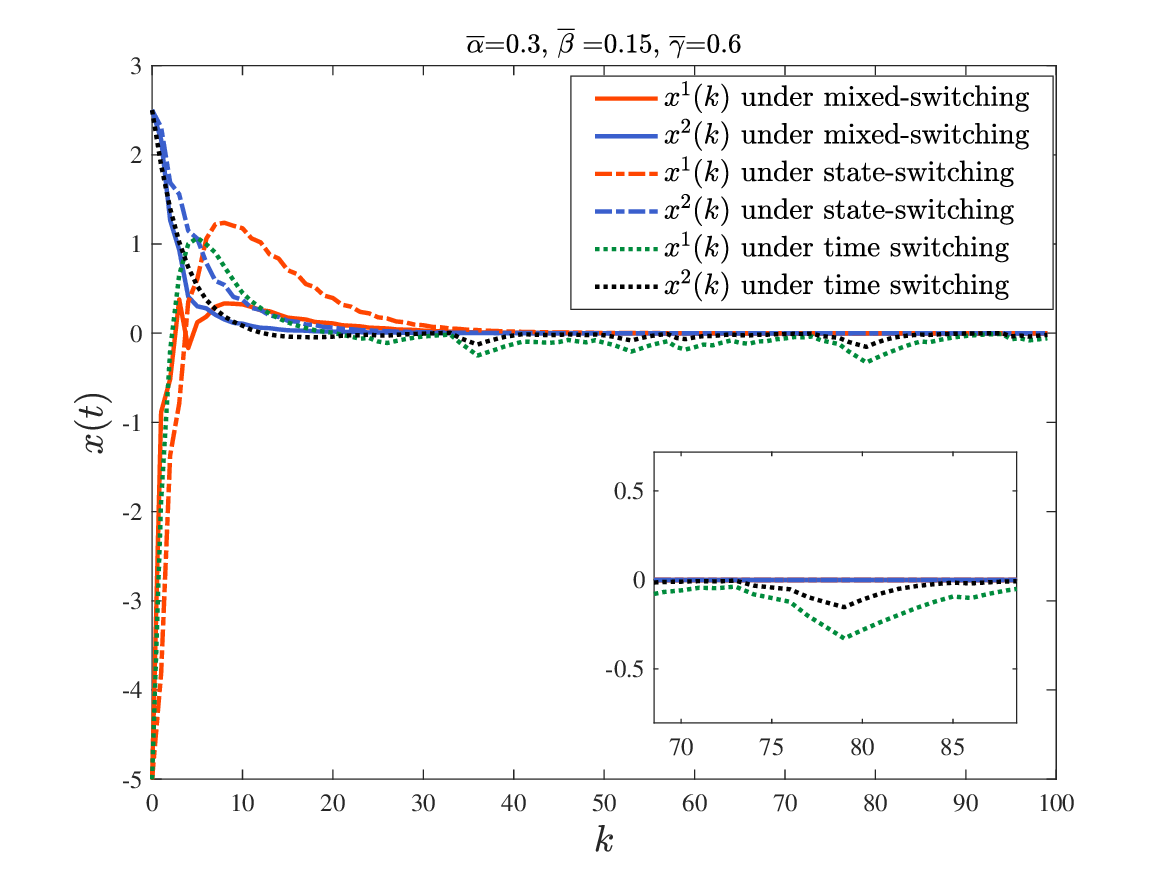}
\vspace{-0.5cm}
\caption{\textcolor{black}{The state trajectories under mixed-switching  for Example 3}}
\label{fig:state1}
\end{minipage}
\end{figure}

When an aircraft executes multi-tasks on one flight, the mode of the aero-engine will switch in order to achieve different control objectives. To obtain an accurate model, the switched modeling technique is used.  Here, we follow the continuous-time switched system parameters in \cite{model}, and then perform the discretization with sampling time being 0.1s. The discrete-time switched system version has the subsystem parameters
\begin{equation*}
\begin{aligned}
		A_1 =& \left[\begin{array}{cc}
	~~~	0.7152 &   0.5893\\
	~~~	0.0051  &  0.7392
	\end{array}\right],~
	B_1 = \left[\begin{array}{c}
		0.0155\\
		0.0044
	\end{array}\right],\\
	A_2 = &\left[\begin{array}{cc}
		~~0.8909    &0.2549\\
		-0.0003  &  0.9233
	\end{array}\right],~
	B_2 = \left[\begin{array}{c}
		0.0186\\
		0.0113
	\end{array}\right],
\end{aligned}
\end{equation*}
For time-dependent switching case, let  $\rho_s = 0.1,~\rho_u =0.4,~\overline{\gamma} = 0.13, ~\alpha = 0.13, ~\beta = 0.13, ~\mu=1.1$. By Theorem \ref{thm2}, the controller gains can be calculated as $	K_1 = \left[\begin{array}{cc}
	-2.4502 &  -1.3115
\end{array}\right],~
K_2 = \left[\begin{array}{cc}
	-4.3778   &-2.5042
\end{array}\right],$
with $\epsilon=  0.0037$, $\tau_d^*  = 7.9036$. The norm of state and the bound $\psi$ are plotted in Fig. \ref{Fig_norm1}.

For the mixed-switching case, choose  parameters $\rho_s = 0.1$, $\lambda =1$, $\mu =2$, $\mu_1 = 1.1$ and $\mu_2 = 1.7$. The controller gains are $	K_1 = \left[\begin{array}{cc}
	-41.4486  &-51.4499
\end{array}\right]$  and $	K_2 = \left[\begin{array}{cc}
	-34.6328 &-32.7794
\end{array}\right].$ 
Fig.  \ref{fig:state1} displays state trajectories under time switching (Theorem \ref{thm2}), mixed-switching (Theorem \ref{thm3}), and state-switching (\ref{state_switching}). It is observed that the system under time switching fails to converge to zero. The performance of the system under mixed-switching and time switching surpasses that of state-switching.

\color{black}

\section{Conclusion}\label{sec_col}

The stabilization for  switched systems in the presence of both DoS attack and deception attack has been investigated.  Sufficient conditions ensuring the \textcolor{black}{practical stability} of the considered system in the mean square sense have been derived where the asynchronous behavior caused by the DoS attack is tackled. The security level has also been explicitly given, which relies on the attack parameters and the initial state. A mixed-switching control strategy has been developed to make sure the system globally asymptotically stable in the presence of DoS attack and deception attack.  In the future, the stealthy deception attack and other types of attacks for switched systems will be discussed.

\appendix

\section{Proof of Theorem \ref{thm1}}\label{app1}
\begin{pf}
	Define \textcolor{black}{a candidate} Lyapunov function
	\begin{equation}\label{equ_Lya}
		V(k) = \widetilde{x}^T(k) P_{\sigma(k)} \widetilde{x}(k).
	\end{equation}
	The proof can be divided into two cases with respect to the dynamics of closed-loop system (\ref{clo-sy1})--(\ref{clo-sy2}): the synchronous stage and the asynchronous stage. Let $		\xi (k)= \left[ 
	\begin{array}{cc}  
		\widetilde{x}^T(k)& x_a^T(k)
	\end{array}
	\right]^T$.
	
	\textbf{Case 1: The synchronous stage}
	
	\textcolor{black}{In this case, the controller mode and the subsystem mode match, i.e., $\sigma(k) = \overline{\sigma}(k)$. And the dynamics of the closed-loop system admits (\ref{clo-sy1})}.
	
	Calculating the difference of $V(k)$ over $[k_s,k_{s+1})$, and let the $p$-th subsystem be activated in this interval, one has
	\begin{align*}
		& \mathbb{E} \left\{\Delta V\left(k\right)| \xi(k)\right\}\\
		=~ &\mathbb{E} \left\{ \widetilde{x}^T(k+1) P_p \widetilde{x}(k+1) -\widetilde{x}^T(k) P_p \widetilde{x}(k)  | \xi(k)\right\}\\
		=~&\widetilde{x}^T(k)\left((\mathcal{A}^1_p)^TP_p\mathcal{A}^1_p+\widetilde{\beta}(\mathcal{A}^2_p)^TP_p\mathcal{A}^2_p-P_p \right.\\
		& +\left(\widetilde{\alpha}\left(1-\overline{\beta}\right)^2+\widetilde{\beta}\left(1-\overline{\alpha}\right)^2+ \widetilde{\alpha}\widetilde{\beta}\right)(\mathcal{A}^3_p)^TP_p \mathcal{A}^3_p \\
		&\left.+2 \widetilde{\beta}\left(1-\overline{\alpha}\right)(\mathcal{A}^2_p)^TP_p\mathcal{A}^3_p \right) \widetilde{x}(k)\\
		& + 2 \widetilde{x}^T(k)\left(	\overline{\alpha}\left(1-\overline{\beta}\right)(\mathcal{A}^1_p)^TP_p \mathcal{A}^4_p\right.\\&
		\left.-\overline{\alpha}\widetilde{\beta} (\mathcal{A}^2_p)^TP_p \mathcal{A}^4_p-\widetilde{\alpha}\left(1-\overline{\beta}\right)^2(\mathcal{A}^3_p)^TP_p \mathcal{A}^4_p\right)x_a(k)\\& +x_a^T(k) \overline{\alpha}\left(1-\overline{\beta}\right)(\mathcal{A}^4_p)^TP_p \mathcal{A}^4_px_a(k)\\
		=~& \xi^T(k) \overline{\Pi}_p  \xi(k)
	\end{align*}
	where $	\overline{\Pi}_p= \left[ 
	\begin{array}{cc}  
		{\Pi}_p^{11}-\rho_s P_p & {\Pi}_p^{12}\\
		*& {\Pi}_p^{22} + \epsilon \textbf{I}
	\end{array}
	\right]
	$.
	
	From (\ref{deception_ass}) and (\ref{con1}), it follows that
		\begin{equation}\label{equ_pf2}
			\begin{aligned}
		&	\mathbb{E}\left\{\Delta V\left(k\right)|\xi (k)\right\} \\
			\leq ~&\mathbb{E}\left\{ \xi^T(k) \overline{\Pi}_p  \xi (k)+ \epsilon \left(\overline{\gamma}^2 - x_a^T(k)x_a (k)\right)\right\}\\
			\leq ~&\mathbb{E}\left\{ - \rho_s \xi^T (k)\text{diag} \left\{ P_p,\textbf{0}\right\}   \xi(k) \right\} + \epsilon \overline{\gamma}^2 \\
			\leq ~&\mathbb{E}\left\{ - \rho_s V\left(k\right) \right\} + \epsilon \overline{\gamma}^2 
			\end{aligned}
	\end{equation}
where the last inequality is derived via (\ref{equ_Lya})

	\textbf{Case 2: The asynchronous stage}
	
	\textcolor{black}{In such a case, the controller mode and the subsystem mode are mismatching, i.e., $\sigma(k)  \neq \overline{\sigma}(k)$. Here, DoS attack successfully occurs i.e., $\beta(k) =1 $ and the system behaves as (\ref{asy-clo}). Then the stage will stop once the DoS attack fails, i.e., $\beta(k) = 0$, and the closed-loop system turns to be (\ref{clo-sy2})}.
	
	In this stage, the $p$-th system is activated in $[k_s,k_{s+1})$ and the $q$-th one is activated in $[k_{s-1},k_{s})$. It is noted that the latest mode will not be transmitted, that is, the DoS attack continuously works. Define the Lyapunov functions with dynamics (\ref{asy-clo}) and (\ref{clo-sy2}) be $V_1(k)$ and $V_2(k)$. $V_1(k)$ and $V_2(k)$ have the same form as $V(k)$, but they represent different system dynamics, which correspond to $\beta(k ) = 1$ and $\beta(k) =0$, respectively.  Thus the overall Lyapunov function can be rewritten as
	\begin{equation}\label{equ_lya2}
	 V\left(k\right) = \beta(k)  V_1\left(k\right) + (1-\beta(k) ) V_2\left(k\right) .
	\end{equation}
Then we have
			\begin{align}
				&\mathbb{E}\{ \Delta V\left(k\right) |\xi(k)\} \label{V_total}\\
				=~&\mathbb{E}\{  \beta(k)\Delta V_1(k)+  (1-\beta(k))\Delta V_2(k)|\xi(k)\} \notag\\
				= ~&\overline{\beta}\mathbb{E}\{\Delta {V}_1\left(k\right)|\xi(k)\} + \left(1-\overline{\beta}\right)\mathbb{E}\{\Delta {V}_2\left(k\right)|\xi(k)\}\notag
			\end{align}
		where $\Delta V_1(k) = V_1(k+1) - V(k)$ and $\Delta V_2(k) = V_2(k+1) - V(k)$ due to the fact that $V(k)$ is determinate for $\xi(k)$.
	The first item in the above formula  can be derived for $k\in[k_s,k_{s+1})$
	\begin{align*}
		\mathbb{E} \{\Delta V_1(k)\textcolor{black}{|\xi(k)}\}=~& \mathbb{E} \left \{ \widetilde{x}^T(k) \left((\overline{\mathcal{A}}^1_{pq})^T P_p \overline{\mathcal{A}}^1_{pq} -  P_p\right) \widetilde{x} (k)\right\}\\
		= ~& \xi^T(k) \overline{\Omega}_{pq} \xi(k)\notag
	\end{align*}
	where $		\overline{\Omega}_{pq}= 
	{\Omega}_{pq}+\rho_u P_p$. 
	Similar to the analysis of (\ref{equ_pf2}), formula (\ref{con2}) yields
		\begin{align}
			\mathbb{E}\left\{\Delta V_1(k)\textcolor{black}{|\xi(k)}\right\} 
			&\leq \mathbb{E}\left\{  \rho_u \xi^T(k) \text{diag} \left\{ P_p,\textbf{0}\right\}   \xi(k) \right\} \notag\\
			&	\leq  \rho_u  \mathbb{E}\left\{V(k) \right\}.\label{V1}
		\end{align}
	For the second item of (\ref{V_total}), one gets
	\begin{align*}
			&\mathbb{E}\{\Delta V_2(k)\textcolor{black}{|\xi(k)}\} \\
			=~&\mathbb{E}\left\{ \widetilde{x}^T(k) \left((\widetilde{\mathcal{A}}^1_{p})^T P_{p}\widetilde{\mathcal{A}}^1_{p}-P_{p}\right. -2 \hat\alpha	(\widetilde{\mathcal{A}}^1_{p})^T P_{p}\widetilde{\mathcal{A}}^2_{p}	\right.\notag\\
			& \left. +\hat\alpha^2 	(\widetilde{\mathcal{A}}^2_{p})^T P_{p}\widetilde{\mathcal{A}}^2_{p}\right) \widetilde{x}(k)+ \alpha^2(k) {x}_a^T(k)(\widetilde{\mathcal{A}}^3_{p})^T P_{p}\widetilde{\mathcal{A}}^3_{p}  {x}_a(k)\notag\\
			&\left.+ \widetilde{x}^T(k)2\alpha(k)\left( (\widetilde{\mathcal{A}}^1_{p})^T P_{p}\widetilde{\mathcal{A}}^3_{p}\right.\left. -\hat \alpha (\widetilde{\mathcal{A}}^2_{p})^T P_{p}\widetilde{\mathcal{A}}^3_{p}\right) {x}_a(k)\textcolor{black}{|\xi(k)}\right\}\notag\\
			=~& \xi^T(k)\overline{\Psi}_{p} \xi(k) \notag
	\end{align*}
	where  $\hat{\alpha} = \alpha(k) - \overline{\alpha}$ and $		\overline{\Psi}_p= \left[ 
	\begin{array}{cc}  
		{\Psi}_p^{11}-\rho_s P_p & {\Psi}_p^{12}\\
		*& {\Psi}_p^{22} + \epsilon \textbf{I}
	\end{array}
	\right]$.
	Hence,  (\ref{con3}) implies that
	\begin{align}
		&\mathbb{E}\left\{\Delta {V}_2\left(k\right)\textcolor{black}{|\xi(k)}\right\} \notag\\
		&\leq \mathbb{E}\left\{ - \rho_s \xi^T\left(k\right) \text{diag} \{ P_p,\textbf{0}\}   \xi\left(k\right) \right\} + \epsilon \overline{\gamma}^2  \notag\\
		&\leq - \rho_s \mathbb{E}\left\{  V\left(k\right) \right\} + \epsilon \overline{\gamma}^2 \label{V2}
	\end{align}
	Substituting (\ref{V1}) and (\ref{V2}) into (\ref{V_total}) provides 
	\begin{align*}
		&\mathbb{E}\{V(k_s+1) \}\\
		=~& \left(\overline{\beta}\overline{\rho}_u+ (1-\overline{\beta})\overline{\rho}_s\right)\mathbb{E}\{ {V}(k_s^+)\}+ (1-\overline{\beta}) \epsilon \overline{\gamma}^2 \\
		\leq~ & \left(\overline{\beta}\overline{\rho}_u + (1-\overline{\beta})\overline{\rho}_s\right)\mathbb{E}\{ {V}(k_s^+)\} +\epsilon \overline{\gamma}^2
	\end{align*}
with $k_s$ being the switching instant and $\overline{\rho}_s = 1-\rho_s$, $\overline{\rho}_u = 1+\rho_u$.

	With the above preparations, for $k \in (k_s,k_{s+1})$, it deduces
	\begin{align}
		&\mathbb{E}\{  V(k) \}\notag \\
		=~&  (1-\overline{\beta}) \sum_{j=0}^{k-k_s-1} \overline{\rho}_s^j\epsilon \overline{\gamma}^2+\Big(\overline{\beta}^{k-k_s} \overline{\rho}_u^{k-k_s}\notag\\
		&+(1-\overline{\beta}) \sum_{i = 0}^{k-k_s-1} \left(\overline{\beta}^{i}\overline{\rho}_u^{i}\overline{\rho}_s^{k-k_s-i}\right)\Big)\mathbb{E}\{ {V}(k_s^+)\} \notag\\
		& + \sum_{i = 1}^{k-k_s-1} \Big(\overline{\beta}^{i} (1-\overline{\beta})\sum_{j=0}^{k-k_s-i-1} \overline{\rho}_s^j  \Big)\epsilon \overline{\gamma}^2\notag\\
		\leq~& g_1(k,k_s) \mathbb{E}\{ {V}(k_s^+)\} + g_2(k,k_s)\epsilon \overline{\gamma}^2\label{int1}
	\end{align}
	where  \\
	\noindent		$g_1(k,k_s) = \overline{\beta}^{k-k_s} \overline{\rho}_u^{k-k_s}  +(1-\overline{\beta})  \sum_{i = 0}^{k-k_s-1} \left(\overline{\beta}^{i}\overline{\rho}_u^{i}\overline{\rho}_s^{k-k_s-i}\right),$
		$g_2(k,k_s) = ~\sum_{j=0}^{k-k_s-1} \overline{\rho}_s^j.$
	It is easy to find that 
$\frac{g_1(k,k_s)}{(\overline{\rho}_s)^{k-k_s}} 
		= \overline{\beta}^{k-k_s} \left({\overline{\rho}_u}\diagup{\overline{\rho}_s}\right)^{k-k_s}
		+(1-\overline{\beta})\frac{1-\overline{\beta}^{k-k_s}\left({\overline{\rho}_u}\diagup{\overline{\rho}_s}\right)^{k-k_s}}{1-\overline{\beta} \overline{\rho}_u \diagup\overline{\rho}_s}
		=c^{k-k_s} + \left(1-\overline{\beta}\right)\frac{1-c^{k-k_s}}{1-c}$
	with $c= \overline{\beta}\frac{\overline{\rho}_u}{\overline{\rho}_s} $.
	Besides, we get 
	$g_2(k,k_s)= ~ \frac{1-\overline{\rho}_s^{k-k_s}}{1-\overline{\rho}_s}
	\leq ~ \frac{1}{\rho_s}$ since $1-\overline{\rho}_s^{k-k_s} \leq 1$ and $1-\overline{\rho}_s = \rho_s$.

	For $k \in (k_s,k_{s+1})$, (\ref{int1}) can be rewritten as
	\begin{align*}
		&\mathbb{E}\{  V(k) \} \\
		\leq~&\Big( c^{k-k_s} + \left(1-\overline{\beta}\right)\frac{1-c^{k-k_s}}{1-c}\Big) \overline{\rho}_s^{k-k_s} \mathbb{E}\{ {V}(k_s^+)\}\\
		& + g_2(k,k_s)\epsilon \overline{\gamma}^2\\
		\leq ~& \Big( c^{k-k_s} + \left(1-\overline{\beta}\right)\frac{1-c^{k-k_s}}{1-c}\Big) \overline{\rho}_s^{k-k_s} \mu \mathbb{E}\{ {V}(k_s^-)\} \\
		&+g_2(k,k_s)\epsilon \overline{\gamma}^2\\
		\leq ~& \frac{2-\overline{\beta}-c}{1-c} \overline{\rho}_s^{k-k_s} \mu \mathbb{E}\{ {V}(k_s^-)\} +g_2(k,k_s)\epsilon \overline{\gamma}^2\\
		\leq ~& \overline{\mu}~ \overline{\rho}_s^{k-k_s} \mathbb{E}\{ {V}(k_s^-)\} +g_2(k,k_s)\epsilon \overline{\gamma}^2.
	\end{align*} 
	\textcolor{black}{The second inequality is derived based on (\ref{con4}), that is, $V(k_s^+) \leq \mu V(k_s^-)$.}
	From (\ref{taud}), it gets $			\overline{\mu} (1-\rho_s)^ {\tau_d}  \leq  1$.
	Hence, we attain
	\begin{equation}\label{lin1}
		\begin{aligned}
			\mathbb{E}\{ V(k_{s}+\tau_d) \}
			\leq \mathbb{E}\{ V(k_{s}^-) \} + g_2(k_{s}+\tau_d, k_{s}) \epsilon \overline{\gamma}^2.
		\end{aligned}
	\end{equation}
	
	Thus, the Lyapunov function admits
	\begin{align}
		&\mathbb{E}\{ V(k_s^-) \}\notag \\
		\leq~ &  \overline{\rho}_s^{k_s-k_{s-1}-\tau_d}	\mathbb{E}\{ V(k_{s-1}^-) \} +g_2(k_s,k_{s-1})\epsilon \overline{\gamma}^2 \notag\\
		\leq~&	\overline{\rho}_s ^ {k_s - s\tau_d} V(0)+g_2(k_s,0)\epsilon \overline{\gamma}^2
	\end{align}
	\textcolor{black}{where $V(0) = \widetilde{x}^T(0) P_{\sigma(0)} \widetilde{x}(0)$.}

	To sum up, for any $k\in [k_s+\tau_d, k_{s+1})$, we always have 		
	\begin{align}\label{lin3}
		~\mathbb{E}\{ V(k) \}	\leq &~ \overline{\rho}_s^{k-k_{s}-\tau_d} \mathbb{E}\{ V(k_{s}^-) \} + g_2(k,k_{s})\epsilon \overline{\gamma}^2\notag\\
		\leq&~	\overline{\rho}_s ^ {k - (s+1)\tau_d} V(0)+g_2(k,0)\epsilon \overline{\gamma}^2\notag	\\
		\leq&~ \overline{\rho}_s ^ {k - (s+1)\tau_d} V(0)+ \frac{1}{\rho_s} \epsilon \overline{\gamma}^2.	
	\end{align}	
	Through similar derivation for $k\in [k_s,k_s+\tau_d)$, it yields
	\begin{align}\label{lin4}
		&\mathbb{E}\{ V(k) \} \notag
		\\ \leq &~ \overline{\mu} ~\overline{\rho}_s^{k-k_s} \mathbb{E}\{ V(k_{s}^-)\} +g_2(k,k_s) \epsilon \overline{\gamma}^2\notag\\ 
		\leq &~ \overline{\mu}~ \overline{\rho}_s^{k- s\tau_d} V(0)+ \overline{\mu} ~\overline{\rho}_s^{k-k_s} g_2(k_s,0)\epsilon \overline{\gamma}^2 +g_2(k,k_s) \epsilon \overline{\gamma}^2\notag\\
		\leq & ~\overline{\mu}~ \overline{\rho}_s^{k- s\tau_d} V(0)+ \overline{\mu} g_2(k,0) \epsilon \overline{\gamma}^2\notag\\
		\leq & ~ \overline{\rho}_s^{k- (s+1)\tau_d} V(0)+ \overline{\mu} \frac{1}{\rho_s} \epsilon \overline{\gamma}^2.
	\end{align}
		Combining (\ref{lin3}) with (\ref{lin4}), one has
	\begin{equation}\label{lin5}
		\begin{aligned}
			\mathbb{E}\{ V(k) \} \leq   ~\overline{\rho}_s^{k- (s+1)\tau_d}V(0) + \overline{\mu} \frac{1}{\rho_s} \epsilon \overline{\gamma}^2.
		\end{aligned}
	\end{equation}
		Since $	V(k) = \widetilde{x}^T(k) P_{\sigma(k)} \widetilde{x}(k)$, it is easy to see  \begin{equation}\label{lin6}
			\min\limits_{p\in\mathcal{M}}\{\lambda_{\min}(P_p)\} \| \widetilde{x}(k)\| ^2 \leq V(k) \leq \max\limits_{p\in\mathcal{M}}\{\lambda_{\max}(P_p)\} \| \widetilde{x}(k)\| ^2.
		\end{equation} 
		Then (\ref{lin5}) can be further expressed by
		$
		\mathbb{E}\{\| \widetilde{x}(k)\| ^2\} \leq   \frac{\overline{\rho}_s^{k- (s+1)\tau_d} \widetilde{x}^T(0) P _{\sigma(0)} \widetilde{x}(0) + \overline{\mu} \frac{1}{\rho_s} \epsilon \overline{\gamma}^2}{	\min\limits_{p\in\mathcal{M}}\{\lambda_{\min}(P_p)\}}.$
		It is clear that condition (11) in Definition \ref{def2} is fulfilled with $\psi = \overline{\mu} \epsilon \overline{\gamma}^2 \diagup (\rho_s\min_{p\in\mathcal{M}}\{\lambda_{\min}(P_p)\})$. Hence, the switched system is \textcolor{black}{practically stable} in the mean square sense.

	Next, we calculate the security level of the switched system under DoS attack and deception attack.
	For any $\varphi > 1$ and $k \in [k_s+\tau_d,k_{s+1})$, it follows that
	\begin{align*}
		&	\mathbb{E}\{\varphi^{k+1}V(k+1)\} - \mathbb{E}\{\varphi^kV(k)\} \\
		\leq ~& \varphi ^{k+1} \mathbb{E}\{\Delta V(k)\textcolor{black}{|\xi(k)}\}+ \varphi^k(\varphi-1) \mathbb{E}\{V(k)\}\\
		\leq ~& \varphi^k\left((\varphi-1)-\varphi \rho_s \right) \mathbb{E}\{V(k) \} + \varphi^{k+1} \epsilon \overline{\gamma}^2.
	\end{align*}
The second inequality is derived based on (\ref{equ_pf2}).
	Selecting $\varphi = \frac{1}{1-\rho_s}$, one thereby gets 
	\begin{align*}
		&\mathbb{E}\{V(k)\}\\ 
		\leq ~&\varphi^{-k+k_s+\tau_d} \mathbb{E}\{V(k_s+\tau_d)\} + (1+\cdots + \varphi^{-k+k_s+\tau_d})\epsilon \overline{\gamma}^2\\
		\leq~ &\varphi^{-k+k_s+\tau_d} \mathbb{E}\{V(k_s+\tau_d)\} + \frac{1-\varphi^{k_s+\tau_d-k}}{1-\varphi^{-1}}\epsilon \overline{\gamma}^2.
	\end{align*}
	In terms of (\ref{lin1}), it attains
	\begin{align*}
		&	\mathbb{E}\{V(k)\}\\ 
		%		\leq & \varphi^{-k+k_s+\tau_d} \mathbb{E}\{V(k_s+\tau_d)\} + \frac{1-\varphi^{k_s+\tau_d-k}}{1-\varphi^{-1}}\epsilon \overline{\gamma}^2\\
		\leq ~& \varphi^{-k+k_s+\tau_d} \mathbb{E}\{V(k_s^-)\} + \frac{1-\varphi^{k_s+\tau_d-k}}{1-\varphi^{-1}}\epsilon \overline{\gamma}^2\\
%		\leq~ & \varphi^{-k+k_{s-1}+2\tau_d} \mathbb{E}\{V(k_{s-1}^-)\} + \frac{1-\varphi^{k_{s-1}+2\tau_d-k}}{1-\varphi^{-1}}\epsilon \overline{\gamma}^2\\
		\leq ~& \cdots\\
		\leq ~& \varphi^{-k+(s+1)\tau_d} V(0) + \frac{1-\varphi^{(s+1)\tau_d-k}}{1-\varphi^{-1}}\epsilon \overline{\gamma}^2\\
		\leq~ & \varphi ^{-k + (s+1)\tau_d} \left( V(0) -\frac{1}{1-\varphi^{-1}}\epsilon \overline{\gamma}^2 \right) +\frac{1}{1-\varphi^{-1}}\epsilon \overline{\gamma}^2 \\
		\leq~ & \max \left\{V(0),\frac{1}{1-\varphi^{-1}}\epsilon \overline{\gamma}^2 \right\}.
	\end{align*}
	Similarly, for $k\in[k_s,k_s+\tau_d)$, we have 
$\mathbb{E}\{V(k)\} \leq ~\overline{\mu} \cdot \mathbb{E}\{V(k_s^-)\}
			\leq \overline{\mu} \cdot \max \left\{V(0),\frac{1}{1-\varphi^{-1}}\epsilon \overline{\gamma}^2 \right\}$.
		From (\ref{lin6}), one has
		$\mathbb{E}\{\| \widetilde{x}(k)\| ^2\} \leq \frac{ \overline{\mu} \cdot \max \left\{ \widetilde{x}^T(0) P _{\sigma(0)} \widetilde{x}(0),\frac{\epsilon \overline{\gamma}^2 }{\rho_s}\right\}}{	\min\limits_{p\in\mathcal{M}}\{\lambda_{\min}(P_p)\}} = \ell$.
	This is the expression of  the security level $\ell$. The proof is completed. 
\end{pf}

\section{Proof of Theorem \ref{thm2}}\label{app2}
\begin{pf}
	Using Schur complement lemma and 
	$X^T Y + Y^T X \leq X^TX + Y^T Y$,
	 (\ref{con1}) can be rewritten as
$		{\overline{\Pi}}_p= \left[ 
		\begin{array}{cc}  
			{\overline{\Pi}}_p^{11}& {\overline{\Pi}}_p^{12}\\
			{\overline{\Pi}}_p^{21}& {\overline{\Pi}}_p^{22}
		\end{array}
		\right] \preceq 0$
	where $		P_p = \text{diag} \{ P_p^1,P_p^2 \}$,
	\begin{align*}
		\overline{\Pi}_p^{11}&=\text{diag}\left\{-P_p^{-1},-P_p^{-1},-P_p^{-1},-P_p^{-1},-P_p^{-1}\right\},\\
		\overline{\Pi}_p^{12} &  = \left[\begin{array}{ccccc}\Theta_{1}^T&\Theta_{2}^T&\Theta_{3}^T&\Theta_{4}^T&\Theta_{5}^T
		\end{array}\right]^T,\\
		\overline{\Pi}_p^{22} &= \text{diag}\left\{-(1-\rho_s)P_p, -\epsilon I^{n_x\times n_x}\right\},\\
		\Theta_{1} &=\left[ \begin{array}{cc}
			\mathcal{A}_p^1 &\overline{\alpha}(1-\overline{\beta})\mathcal{A}_p^4
		\end{array}\right],\\
		\Theta_{2} &= \left[\begin{array}{cc}
			\sqrt{\overline{\alpha}\widetilde{\beta}} \mathcal{A}_p^2 & -\sqrt{\overline{\alpha}\widetilde{\beta}} \mathcal{A}_p^4
		\end{array}\right],\\
		\Theta_{3} &=\left[ \begin{array}{cc}
			\sqrt{\widetilde{\beta}(2-2\overline{\alpha})} \mathcal{A}_p^2 & 0
		\end{array}\right],\\
		\Theta_{4} &= \left[\begin{array}{cc}
			\sqrt{\widetilde{\alpha}}(1-\overline{\beta}) \mathcal{A}_p^3 & -\sqrt{\widetilde{\alpha}}(1-\overline{\beta}) \mathcal{A}_p^4
		\end{array}\right],\\
		\Theta_{5} &= \left[\begin{array}{cc}
			\sqrt{ \widetilde{\beta}(1-\overline{\alpha})(2-\overline{\alpha})} \mathcal{A}_p^3 & 0
		\end{array}\right].
	\end{align*}
	Pre- and post-multiplying the above inequality by
	$\text{diag} \{ Q_p, Q_p,Q_p,Q_p,Q_p, I^{3n_x\times 3n_x} \}$
	and its transpose with
	$	Q_p = \text{diag} \{ \Xi_p E_p, P^2_p\}$,~
	$\Xi_p= \left[
	\begin{array}{cc}
		\Xi_p^{11}&\Xi_p^{11}\\
		0^{(n_x-p) \times p} &\Xi_p^{11}
	\end{array}
	\right]$,~
	$\Xi_p^{11} \in \mathbb{R}^{p \times p}$,~$\Xi_p^{12} \in \mathbb{R}^{p \times (n_x-p)}$, $\Xi_p^{22} \in \mathbb{R}^{(n_x-p) \times (n_x-p)}$, 
	and using the fact
	$X +X^T - X Y^{-1} X^T  - Y \leq -(Y-X) Y^{-1} (Y-X)^T \leq 0,$ (\ref{conn1}) follows directly with $R_p =\Xi_p E_p B_pK_p = \Xi_p^{11} K_p$.
	
	Repeating the above steps, we can get (\ref{conn3}) from (\ref{con3}). Moreover, (\ref{con2}) is equivalent to
	$\left[
	\begin{aligned}
		\begin{array}{cc}
			-P_p^{-1}& \overline{\mathcal{A}}_{pq}^{1} \\
			* & -(1+\rho_u) P_p
		\end{array}
	\end{aligned}
	\right]\preceq 0.$
	Then pre- and post-multiplying the inequality by 
	$	\text{diag} \{ Q_{pq}, I \}$ and its transpose with $Q_{pq} = \text{diag} \{ \Xi_q E_p, P^2_p\}$, (\ref{conn2}) is obtained. This completes the proof.
\end{pf}

\section{Proof of Theorem \ref{thm3}}\label{app3}
\begin{pf}
	When controller (\ref{cont2_1}) is activated,  define the Lyapunov-like function as 
	\begin{equation}
		V(k) = {x}^T(k) P_{\sigma(k)} {x}(k),
	\end{equation}
	then we have
	\begin{align*}
		&\mathbb{E} \left\{\Delta V\left(k\right)| \xi(k)\right\}\\
		=~&\mathbb{E} \left\{{x}^T(k+1) P_p {x}(k+1)-{x}^T(k) P_p {x}(k)| \xi(k)\right\}\\
		=~& \xi^T(k) {\varGamma}_p  \xi(k)
	\end{align*}
	where $\xi(k) = \left[\begin{array}{c}
		x(k) \\
		x_a(k)
	\end{array}\right]
	$, $	\varGamma_p =~ \left[\begin{array}{cc}
		\varGamma_p^{11}&  \varGamma_p^{12}\\
		*	& \varGamma_p^{22}
	\end{array}
	\right]$,
	\begin{align*}
		\varGamma_p^{11} =~& A^T_p P_p A_p-P_p + 2(1-\overline{\alpha})(1-\overline{\beta}) A^T_p P_p B_pK_p \\
		&+(1-\overline{\alpha})(1-\overline{\beta})  (B_pK_p)^TP_pB_pK_p,\\
		\varGamma_p^{12} 	=~& \overline{\alpha}(1-\overline{\beta}){{A}}^T_pP_p B_pK_p,\\
		\varGamma_p^{22} =~&	\overline{\alpha}\left(1-\overline{\beta}\right)(B_pK_p)^TP_p B_pK_p.
	\end{align*}
	Due to $\varLambda_p \preceq 0$, one has
$	\left[\begin{array}{cc}
			{\varGamma}_p^{11}+\rho_sP_p&  {\varGamma}_p^{12}\\
			*	& {\varGamma}_p^{22}-\epsilon \textbf{I}
		\end{array}
		\right]\preceq 0$
	which yields
$	\mathbb{E}\left\{\Delta V\left(k\right)|\xi (k)\right\} 
		\leq \mathbb{E}\left\{ - \rho_s V\left(k\right) \right\} + \epsilon \overline{\gamma}^2 $.
	
	From (\ref{thm4_2}), it follows that 
$	V(k_s^+) \leq \mu_1 V(k_s^-)$
with the fact that $\sigma(k_s^-) = q $ and $\sigma(k_s^+) = p$.
Using  a similar analysis as that in Theorem \ref{thm1}, we have
	\begin{equation}
	\begin{aligned}
		\mathbb{E}\{ V(k_{s}+\tau_{d1}) \}
		\leq \mathbb{E}\{ V(k_{s}^-) \} + g_2(k_{s}+\tau_{d1}, k_{s}) \epsilon \overline{\gamma}^2
	\end{aligned}
\end{equation}
where $	g_2(k_1,k_2) = ~\sum_{j=0}^{k_1-k_2-1}(1-\rho_s)^j$.

	Similarly,	define the Lyapunov-like function for \textbf{\textit{Strategy 2}} as 
$		V(k) = {x}^T(k) Q_{\sigma(k)} {x}(k)$.
	From (\ref{clo_3}) and  (\ref{thm4_3}), it is derived that
	\begin{align*}
			\mathbb{E} \left\{\Delta V\left(k\right)| \xi(k)\right\}
		=~& x^T(k)(A_p^TQ_p A_p-Q_p) x(k)\\
	\leq~&-x^T(k)\lambda (\mu Q_q-Q_p) x(k).
	\end{align*}
	In the \textbf{\textit{Strategy 2}}, the closed-loop system is determined and the expectation is omitted for brief.
	For any $k\in [k_s,k_s+\tau_{d2})$,  we have $V(k) \leq (1+\lambda)^{k-k_s}V(k_s)$ since $V(k+1) \leq~V(k)+x^T(k)\lambda Q_p x(k)-x^T(k)\mu \lambda Q_qx(k) \leq V(k)+x^T(k)\lambda Q_p x(k) = (1+\lambda)V(k)$.
	Thus one has $V(k_s+\tau_{d2}) \leq (1+\lambda)^{\tau_{d2}}V(k_s^+)$. Moreover, it is found that $\mu V(k_s^+) < V(k_s^-)$ from switching condition (\ref{switching_law2}) where $k_s^+$ and $k_s^-$ denote the time when the system switches from the previous subsystem to the next subsystem.
	Applying the switching strategy $\tau_{d2} \leq \tau_{d2}^* =\frac{\ln \mu}{\ln (1+\lambda)}$, it yields
$			V(k_s+\tau_{d2}) \leq (1+\lambda)^{\tau_{d2}}V(k_s^+)\leq (1+\lambda)^{\tau_{d2}}V(k_s^-)/\mu
			 \leq V(k_s^-).$

	For any  $k\in[k_s+\tau_{d2},k_{s+1})$,	according to the switching strategy, one always gets $\Delta V(k) \leq 0$ since $x^T(k)(\mu Q_q-Q_p)x(k) \geq 0$ from the switching condition (\ref{state_switching}). Therefore, $V(k) \leq V(k_s+\tau_{d2})$. Thus, it is deduced that 
$		V(k_{s+1}^-) \leq 	V(k_{s}^-)$.

Next, we consider the Lyapunov jump when there is a switching between two strategies.
	Let $k_a$ and $k_b$ be the instants where control method switches from (\ref{cont2_1}) to (\ref{cont2_2}) and that switches from (\ref{cont2_2}) to (\ref{cont2_1}). Hence, (\ref{thm4_4}) and (\ref{thm4_5}) imply that 
	\begin{align}
		\mu	V(k_a^+) \leq V(k_a^-),\label{equ_thm4_3}\\
		\mu_2	V(k_b^-) \geq V(k_b^+).\label{equ_thm4_4}
	\end{align}

	Suppose that the next switching instant is $k_s$ after $k_a$ and the latest switching instant is $k_{s-1}$, i.e., $k_{s-1} \leq k_a \leq k_s$. Thus we have $\mathbb{E}\{V(k_s^-)\}\leq\mathbb{E}\{V(k_a^-)\}\leq \mathbb{E}\{V(k_{s-1}^-)\}$  following  \textbf{\textit{Strategy 2}} and (\ref{equ_thm4_3}). Similarly, let $k_{\overline{s}-1}\leq k_b \leq  k_{\overline{s}}$, then $\mathbb{E}\{V(k_{\overline{s}-1}^-)\}\geq \mathbb{E}\{V(k_b^-)\} \geq \mathbb{E}\{V(k_{\overline{s}}^-)\}$ by considering (\ref{equ_thm4_4}) and $\tau_{d1} \geq -\frac{\ln \mu_2}{\ln (1-\rho_s)}$.
	 Using (\ref{equ_thm4_4}), switching law (\ref{switching_law}) and controller switching condition, one obtains $\mathbb{E}\{V(k_a)\} \geq \mathbb{E}\{V(k_b)\}$ where $k_a < k_b$. 
	
	To sum up, it is found that $\lim _{t\rightarrow \infty} \mathbb{E}\{V(t)\} \rightarrow 0$, since the Lyapunov-like functions are non-increasing in each interval by comparing the starting instant and the ending instant of each switching interval. Therefore, the state will converge to the equilibrium and  the closed-loop switched system is globally asymptotically stable in the sense mean square sense.	\qed
\end{pf}

\bibliographystyle{apalike}        % Include this if you use bibtex 
\bibColoredItems{black}{LMI,F18,model,DC,ss,AE-5,AE-1,AE-2,AE-3,AE-7,AE-8,Lin,asy1,asy3}
\bibliography{mybibfile}           % and a bib file to produce the 

\end{document}